\documentclass[11pt]{article}

\usepackage{arxiv}

\usepackage[utf8]{inputenc} 
\usepackage[T1]{fontenc}    
\usepackage{url}            
\usepackage{booktabs}       
\usepackage{amsthm,amsmath,amsfonts}
\usepackage{natbib}
\usepackage{float}
\usepackage{subcaption}
\usepackage{algorithm2e}
\usepackage{xcolor}
\usepackage{multirow}
\usepackage{color,soul}
\usepackage{blkarray}
\usepackage{nicefrac}
\usepackage[colorlinks,citecolor=blue,urlcolor=blue,filecolor=blue,backref=page]{hyperref}
\usepackage{microtype}      
\usepackage{lipsum}
\usepackage{graphicx}
\usepackage{anyfontsize}
\usepackage{afterpage}
\usepackage{adjustbox}
\usepackage{color, colortbl}
\usepackage{mathbbol}
\usepackage{mathrsfs}
\usepackage{hhline}
\usepackage{rotating}
\usepackage{comment}
\usepackage{tabularx}
\usepackage{sectsty}
\usepackage{amsmath}
\usepackage{booktabs}
\usepackage{amssymb}
\usepackage{multirow}
\usepackage{subcaption}
\usepackage{bm}
\usepackage{float}
\usepackage{array}
\usepackage{lscape}
\usepackage{caption}
\usepackage{xcolor}

\definecolor{Gray}{gray}{0.9}

\allowdisplaybreaks

\usepackage{arydshln}
\makeatletter
\renewcommand*\env@matrix[1][*\c@MaxMatrixCols c]{%
	\hskip -\arraycolsep
	\let\@ifnextchar\new@ifnextchar
	\array{#1}}
\makeatother

\RequirePackage{doi}

\usepackage{bbm} 

\newcommand{\tCollapsedurl}{\if0\blind
	{
		\url{\tCollapsedurl}
	} \fi
	\if1\blind
	{
		\url{github.com/}\texttt{[url redacted in blinded version]}
	} \fi}

\title{Nonhomogeneous hidden semi-Markov models for toroidal data}

\author{
Francesco Lagona\\
    \scriptsize{Dpt. of Political Sciences}\\
    \scriptsize{Roma Tre University}\\
    \scriptsize{\texttt{francesco.lagona@uniroma3.it}}
\And
    Marco Mingione\\
    \scriptsize{Dpt. of Political Sciences}\\
    \scriptsize{Roma Tre University}\\
    \scriptsize{\texttt{marco.mingione@uniroma3.it}}
}

\begin{document}

\RestyleAlgo{boxruled}

	\def\spacingset#1{\renewcommand{\baselinestretch}%
		{#1}\small\normalsize} \spacingset{1}

\maketitle
\begin{abstract}
A nonhomogeneous hidden semi-Markov model is proposed to segment toroidal time series according to a finite number of latent regimes and, simultaneously, estimate the influence of time-varying covariates on the process' survival under each regime. The model is a mixture of toroidal densities, whose parameters depend on the evolution of a semi-Markov chain, which is in turn modulated by time-varying covariates through a proportional hazards assumption. Parameter estimates are obtained using an EM algorithm that relies on an efficient augmentation of the latent process. The proposal is illustrated on a time series of wind and wave directions recorded during winter.
\end{abstract}

\section{Introduction}\label{intro}
Pairs of circular observations are often referred to as toroidal data because they can be represented as points on a torus, the cartesian product of two circles. Toroidal data arise in numerous contexts. Examples include pairs of wind and wave directions \citep{Lagona2013},  earthquake data consisting of the pre-earthquake direction of steepest descent and the direction of lateral ground movement \citep{Rivest1997}, and peak systolic blood pressure times, converted to angles, during two separate time periods \citep{fisher1983correlation}. Additional notable applications arise from computational biology as either protein backbone conformational angles \citep{Lennox2009density} or phase angles of circadian-related genes in two tissues \citep{Liu2006phase} or orthologous genes shared by circular prokaryotic genomes \citep{Shieh2011modeling}.

The statistical analysis of toroidal data is inherently different from the traditional analysis of bivariate continuous data, due to the wraparound nature of their domain and the difficulties in modelling the dependence between two angular measurements (circular correlation). Luckily, the spread of toroidal data across multiple disciplines has pushed the literature towards the definition of several distributions on the torus (see the reviews by \citet{Ley2017} - Sections 2.4-2.5 - and by \citet{pewsey_recent_2021} - Section 3.2) and a whole library of toroidal distributions is nowadays available. As a result, when the data are in the form of a sequence of independent and identically distributed observations, they can be efficiently analyzed by fitting one of these distributions.

In most case studies, however, toroidal data are heterogeneous and dependent across space and/or time, motivating models where toroidal densities are just a building block among other model components. For example, the data that motivated this study are in the form of a time series of wind and wave directions, recorded by a buoy at regular time intervals. Not only are these data heterogeneous with a distribution that dynamically varies across different latent sea regimes, but the duration of these regimes (dwell times) may vary according to time-varying weather conditions (e.g., wind speed), introducing complex auto-correlation structures.

This paper introduces a non-homogeneous hidden semi-Markov model (HSMM) that accounts for circular correlation and heterogeneity, simultaneously allowing for flexible dwell times. Under this model, the distribution of the data is approximated by a mixture of toroidal densities, whose parameters depend on the evolution of a latent, nonhomogeneous semi-Markov chain. While the toroidal density accommodates circular correlation, the mixture controls for heterogeneity and, finally, the semi-Markov chain allows flexible dwell times whose distribution is modulated by time-varying covariates and it is therefore non-homogeneous.

The general idea of modelling time series by finite mixtures with time-varying parameters is not new. When the latent process that drives these parameters is a Markov chain, the mixture is referred to as a hidden Markov model (HMM). HMMs are a natural extension of mixture models to the temporal setting. Under a HMM, observations are segmented according to the latent state that is conditionally expected each time, given the observed data. As a result, while under a mixture model the data segmentation relies on similarities in the variables space, under a HMM it is also fueled by similarities that occur in a temporal neighbourhood. The popularity of HMMs stems from the intuitively appealing and convenient Markov hypothesis: if the past and the present of the chain are known, then the future evolution of the chain is determined only by its present state, or equivalently, the past and the future are conditionally independent given the present (state). Under this setting, the past history of the chain plays no role in its future evolution, often referred to as the memoryless property of a Markov process. HMMs for toroidal time series \citep{Lagona2013} or, more generally, multivariate time series with mixed linear and circular components \citep{bulla2012multivariate} have already been proposed in the literature.

Under a HMM, however, the distribution of the sojourn time of each state is geometric, which is a consequence of the memoryless property of the latent process. HSMMs are instead driven by a semi-Markov chain that relaxes the memoryless property by allowing dwell time distributions that are not necessarily geometric. In this way, under a HSMM, data segmentation is integrated with flexible dwell time modelling. This can be helpful in studies when the interest is not only in the identification of the regimes that characterize a time series but also in the expected duration of each regime.

Although HSMMs have not yet been exploited for toroidal time series, they are gaining popularity in environmental studies. Recent examples include GPS tracking data of animal movements \citep{Pohle2022}, multivariate pollutant concentrations \citep{Merlo2022quantile}, migratory bird count data \citep{nicol2023flywaynet} and regime shifts in ocean density variability \citep{economou2019hidden}. These proposals however rely on homogeneous dwell time distributions. While homogeneity can be a realistic assumption in selected applications, it becomes a major drawback when a study aims at estimating the influence of time-varying covariates on the dwell time distribution. We achieve this goal by modelling dwell times by regime-specific proportional hazards (PH) regressions where time-varying covariates tune the hazard of a transition to a different regime. PH methods have been recently proposed by \citet{benny2023} as an intuitively appealing strategy to model transition probabilities of two-state Markov chains under a HMM framework and our proposal can be therefore viewed as an extension of this idea to the HSMM setting.

Despite its generality, the proposed model can be efficiently estimated by an EM algorithm that relies on the well-known representation of a semi-Markov chain as an augmented Markov chain where the information about the state of the chain is complemented by the dwell time spent in that state \citep{Anselone1960}. This representation has been already exploited by \citet{Langrock2011} in a HSMM context. We extend this approach to an EM algorithm that fully integrates PH regressions in the distribution of the latent semi-Markov chain.

The rest of the paper is organized as follows. The marine data that motivated this work are described in Section \ref{sec:data}, while the model and the estimation procedures are respectively described in Sections \ref{sec:model}, \ref{sec:estimation} and \ref{sec:comp.details}. The proposal is first tested on simulated data (Section \ref{sec:simulation}) and then exploited to segment the marine data that motivated this study (Section \ref{sec:application}). Section \ref{sec:discussion} finally summarizes relevant discussion points. The whole code to reproduce both the simulation experiments and the results on the real data are publicly available in a GitHub repository accessible at \url{https://github.com/minmar94/HSMM_covariates}.

\section{Wind and wave direction}\label{sec:data}
Time series of wind and wave directions are routinely collected by  environmental agencies to identify sea regimes, that is, typical distributions that these data take under specific environmental conditions. Detecting environmental regimes is crucial across several problems of marine research. Recent applications include studies of the drift of floating objects and oil spills \citep{gu2024_floating}, the design of marine structures \citep{fang2022_bridges} and the analysis of coastal erosion \citep{flor-blanco2021_coastal}.

\begin{figure}[ht]
\centering
\begin{subfigure}[b]{.3\textwidth}
    \centering
    \includegraphics[width = .95\textwidth]{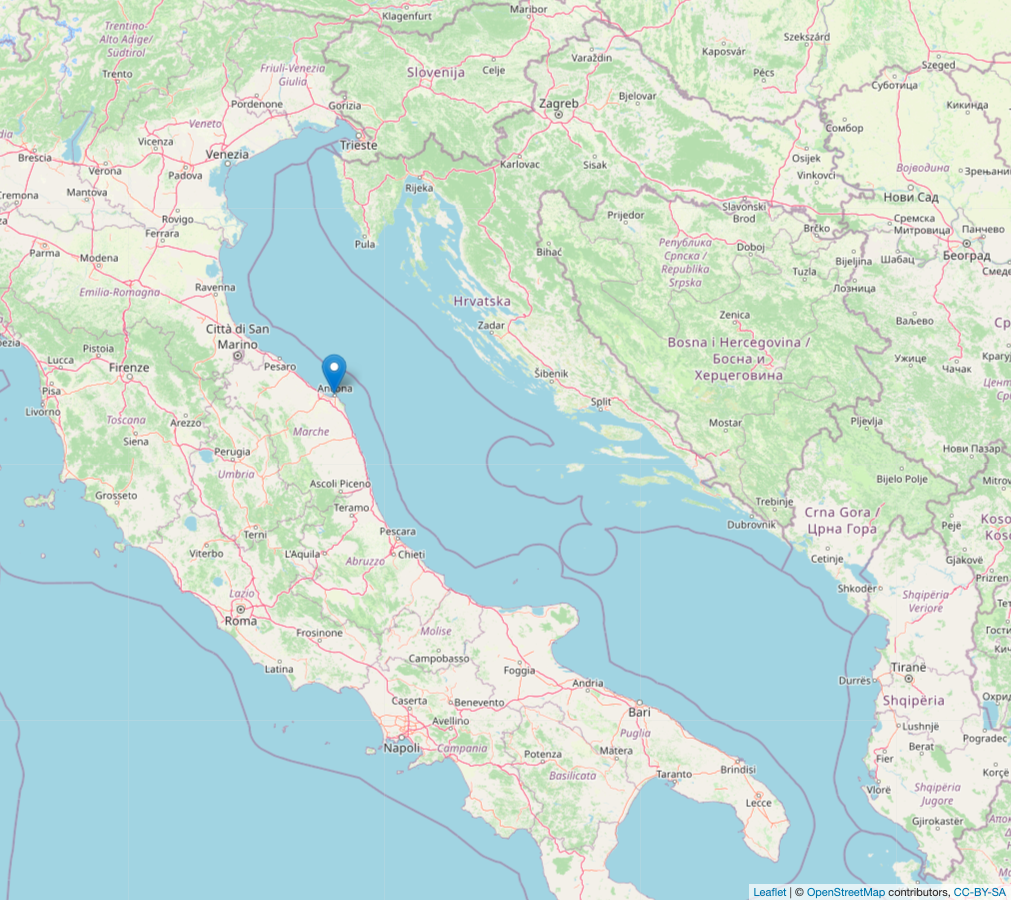}
    \caption{}
    \label{fig:buoyloc}
\end{subfigure}
\begin{subfigure}[b]{.3\textwidth}
    \centering
    \includegraphics[width = .8\textwidth]{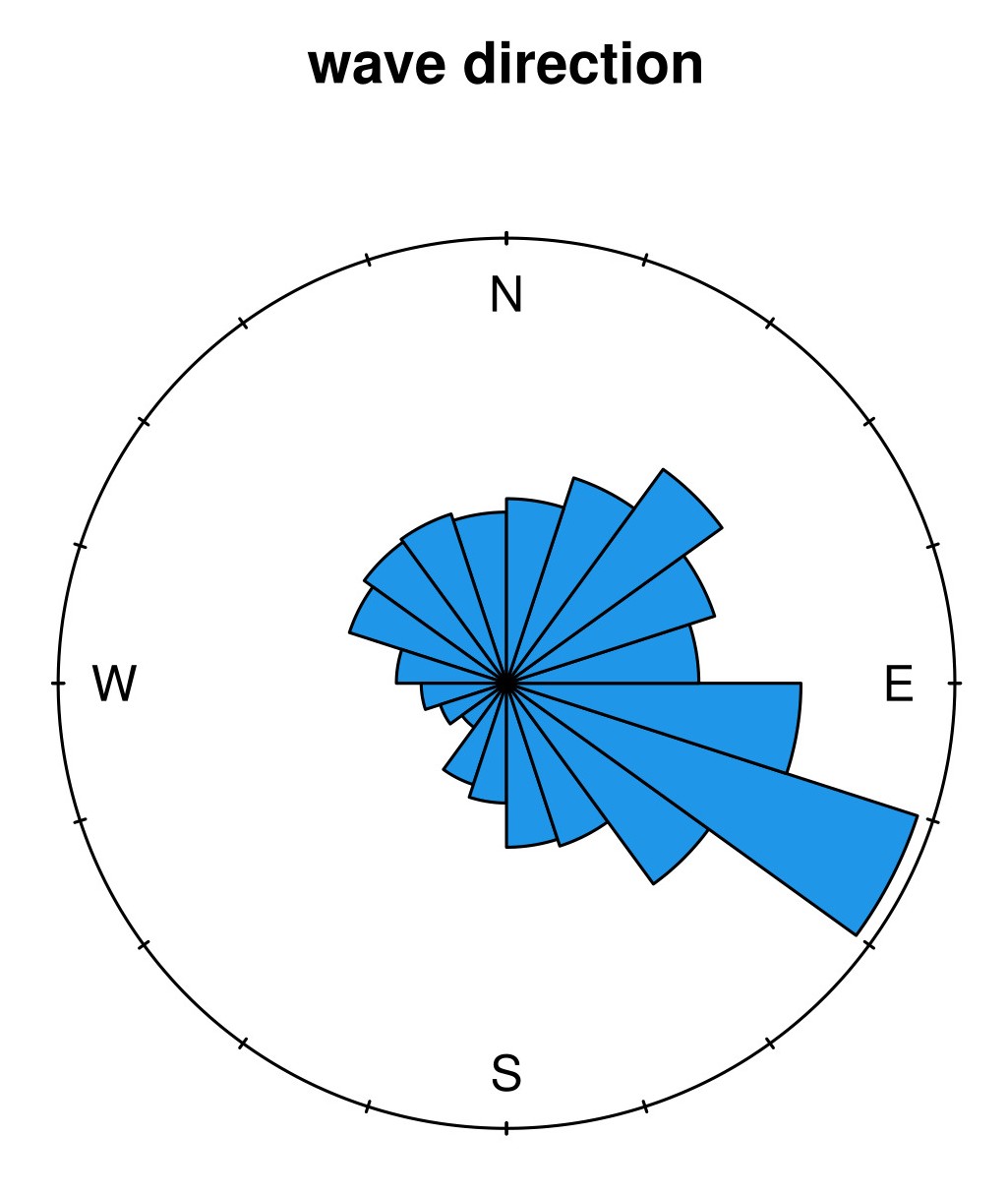}
     \caption{}
    \label{fig:obswavedir}
\end{subfigure}
\begin{subfigure}[b]{.3\textwidth}
    \centering
   \includegraphics[width = .8\textwidth]{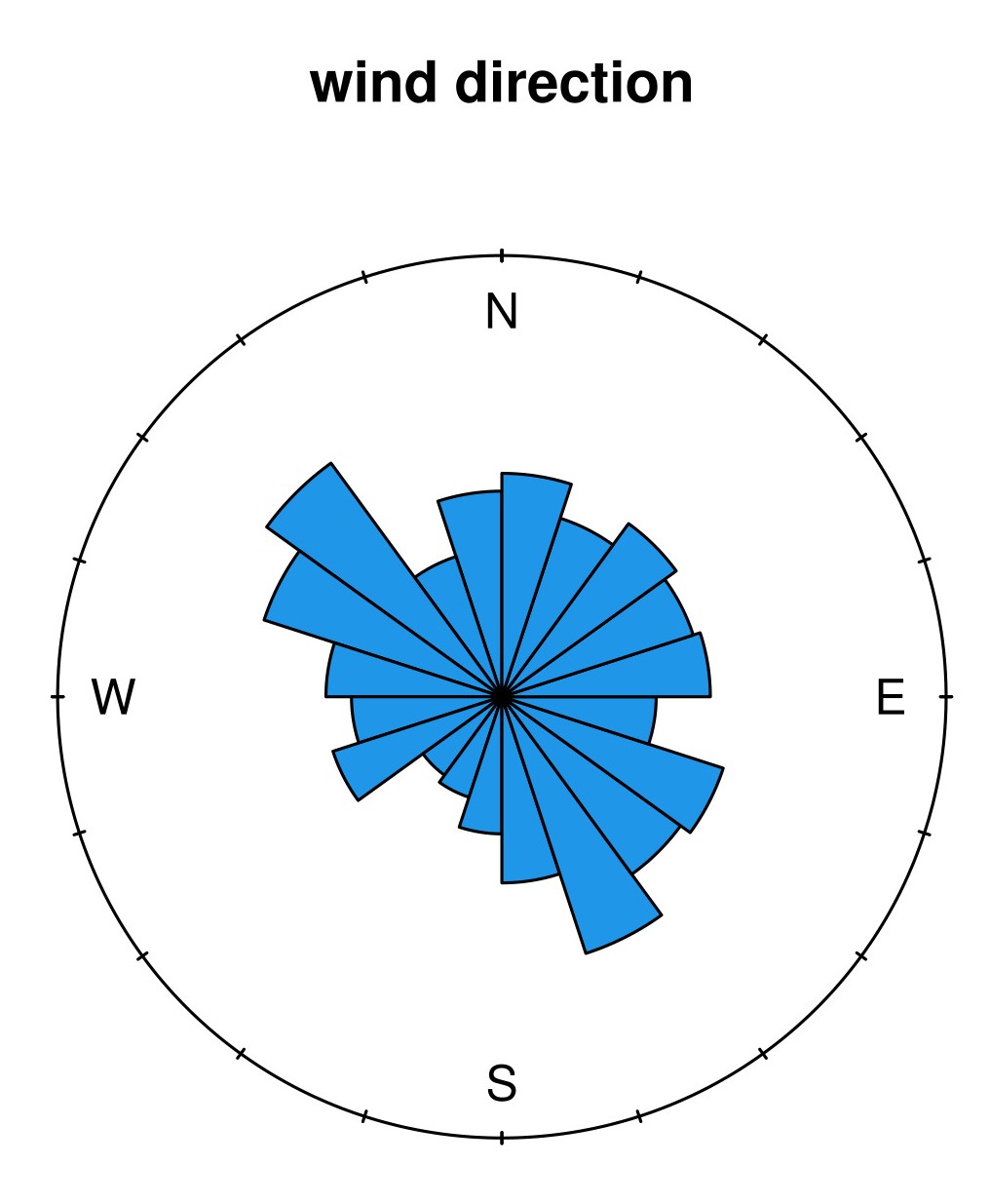}
     \caption{}
    \label{fig:obswinddir}
\end{subfigure}
\caption{(a) The Adriatic sea and the location of the Ancona buoy (Latitude $43^\circ 37^\prime 29.16^{\prime \prime}$ N); Longitude
$13^\circ 30^\prime 23.46^{\prime \prime}$ E); (b) observed wave directions and (c) wind directions.} \label{fig:ancona_buoy}
\end{figure}

\begin{figure}[ht]
    \centering
        \begin{subfigure}[b]{.48\textwidth}
        \centering
            \includegraphics[width = .9\textwidth]{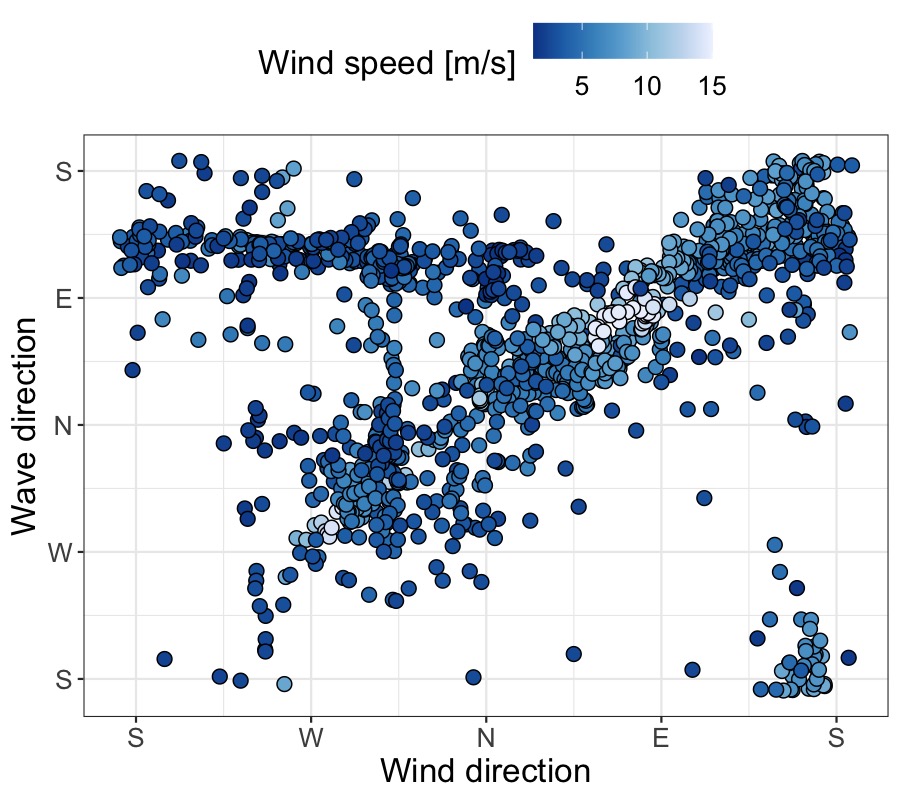}
        \caption{}
        \label{fig:obs2D}
        \end{subfigure}
        \begin{subfigure}[b]{.48\textwidth}
        \centering
            \includegraphics[width = .8\textwidth]{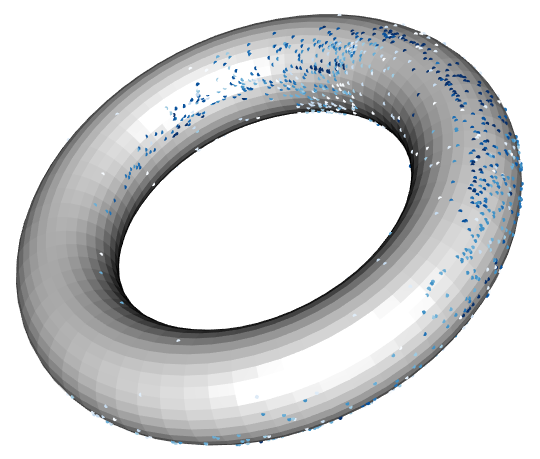}
        \caption{}
        \label{fig:obs3D}
        \end{subfigure}
    \caption{(a) Joint distribution of wave and wind directions, projected to a plane obtained by unwrapping a torus; (b) the same data displayed in their original manifold. Dots are coloured according to contemporaneous wind speed [m/s].}
    \label{fig:data}
\end{figure}

The data that motivated this work are in the form of a time series of $T=1326$ semi-hourly wave and wind directions, recorded between February 15th and March 16th 2010 by the Ancona buoy, located in the Adriatic Sea at about 30 km from the coast (Figure \ref{fig:buoyloc}). Precisely, the data are gathered in the form of angles that indicate the average direction {\it from} which the wave travels and the wind blows, during a period of 30 minutes. Figures \ref{fig:obswavedir} and \ref{fig:obswinddir} show the two marginal distributions of wind and wave directions, in the form of rose diagrams. The data joint distribution is instead shown by Figure \ref{fig:data} in the form of two scatterplots, one with dots displayed on the original toroidal manifold (Figure \ref{fig:obs3D}), and the other one with dots projected on the plane that is obtained by unwrapping the torus (Figure \ref{fig:obs2D}). Dots are filled by colors according to wind speed, a covariate recorded by the buoy every 30 minutes.

Some features of the data can be straightforwardly interpreted by recalling the meteorology of the sea that surrounds the Ancona buoy.
In wintertime, waves in the Adriatic Sea are typically generated by the southeastern Sirocco wind and the northern
Bora wind. These conditions can be associated with the two modes of the rose diagram in Figure \ref{fig:obswavedir}. Sirocco arises from a warm, dry,
tropical air mass that is pulled northward by low-pressure cells moving eastward across the Mediterranean Sea. By contrast, Bora episodes
occur when a polar high-pressure area sits over the snow-covered mountains of the interior plateau behind the coastal mountain range and a calm low-pressure area lies further south over the warmer Adriatic.

Remarkably, wind and wave directions are not always synchronized, as shown by the data scatterplot (Fig. \ref{fig:obs2D}). In the open sea, waves can travel freely, without being obstructed by physical obstacles, such as coastlines. As a result, the wind energy is fully transferred to the sea surface and wind and wave directions are highly correlated. In a semi-enclosed basin like the Adriatic Sea, instead, wave direction is modulated by the orography of the area and, as a result, wind and wave directions are not necessarily highly correlated. Orographic effects are often held responsible for the inaccuracy of numerical wave models in the Adriatic Sea \citep{Bertotti2009}, motivating statistical methods that segment these data into a small number of latent classes, conditionally on which the distribution of the data takes a shape that is easier to interpret than the shape taken by the marginal distribution. Figure \ref{fig:obs2D}, where observations are coloured according to contemporaneous wind speeds, not only offers first-hand evidence of the presence of at least two clusters. It also shows that high values of wind speed tend to appear more often within one of these classes: a clear-cut clue of the influence of wind speed on data heterogeneity.

By modelling these data by a mixture of toroidal densities with parameters driven by a non-homogeneous semi-Markov chain, we aim to identify meaningful environmental regimes, simultaneously accomplishing two further goals. First, we estimate the distribution of the dwell time spent by the observed process in each regime. Regimes are therefore not only classified according to different shapes of the data distribution, as it is typically done in model-based segmentation, but also according to the distribution of the associated dwell times. Second, we estimate the influence of environmental time-varying covariates (e.g., wind speed) on the duration of these regimes, hence avoiding unrealistic assumptions of homogeneous conditions under which the observed process evolves.

\section{A toroidal hidden semi-Markov model}\label{sec:model}
Let $\bm{y}=(\bm{y}_t, t=1, \ldots T)$ be a bivariate time series, where $\bm{y}_t=(y_{t1}, y_{t2})$ is a vector of two circular observations $-\pi < y_{t1},y_{t2}\leq \pi.$ Further, let $\bm{u}=(\bm{u}_t, t= 1, \ldots T)$ be a sequence of latent multinomial random variables $\bm{u}_t=(u_{t1} \ldots u_{tK})$ with one trial and $K$ classes (or regimes), whose binary components represent class membership at time $t$. Our proposal is a hierarchical model where the joint distribution of the time series is obtained by integrating a parametric conditional distribution $f(\bm{y} \mid \bm{u};\bm{\theta})$ of the observed data given the latent classes (the observation process) with respect to the parametric distribution $p(\bm{u};\bm{\eta})$ of the latent classes (the latent process), namely
\begin{equation} \label{eq:hierachical_model}
    f(\bm{y}; \bm{\theta}, \bm{\eta})=\sum_{\bm{u}}f(\bm{y} \mid \bm{u}; \bm{\theta})p(\bm{u}; \bm{\eta}).
\end{equation}

\subsection{The observation process}
We assume that the toroidal observations are conditionally independent given the latent classes. Under this setting, the observation process is driven by a family of $K$ toroidal densities $f_k(\bm{y})=f(\bm{y}; \bm{\theta}_k)$, which represent the conditional distribution of the data under each regime and are known up to an array $\bm{\theta}=(\bm{\theta}_1,\ldots, \bm{\theta}_K)$ of parameters, namely
\begin{equation}f(\bm{y} \mid \bm{u}; \bm{\theta})=\prod_{t=1}^{T}\prod_{k=1}^{K}f(\bm{y}_t; \bm{\theta}_k)^{u_{tk}}\label{eq:conditional_data_distribution}.\end{equation}
\begin{figure}
     \centering
    \begin{subfigure}[b]{.48\textwidth}
    \centering
    \includegraphics[width=.8\textwidth]{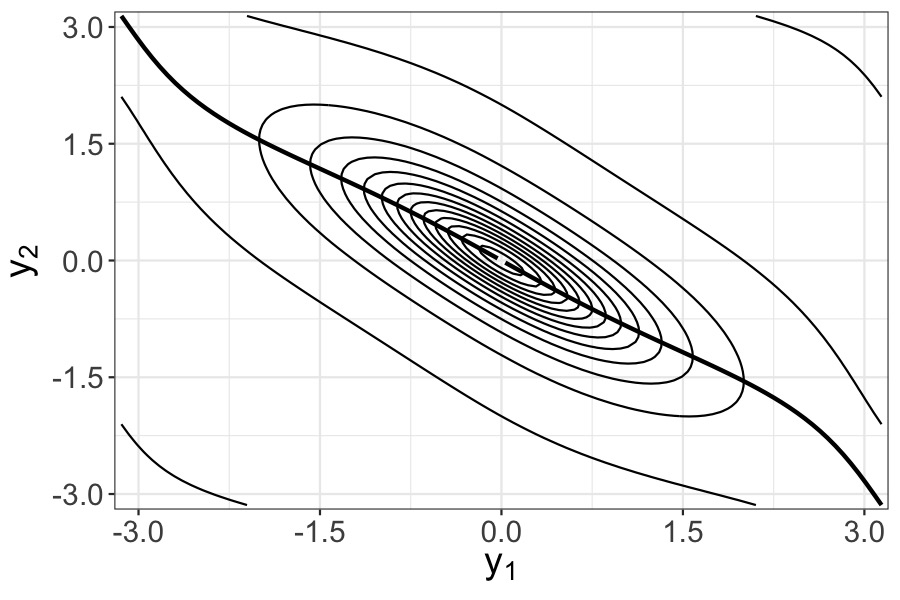}
    \caption{$\kappa_1 = \kappa_2 = 0.3$, $\rho = -0.4$.}
    \label{fig:circ1}
    \end{subfigure}
     \begin{subfigure}[b]{.48\textwidth}
    \centering
    \includegraphics[width=.8\textwidth]{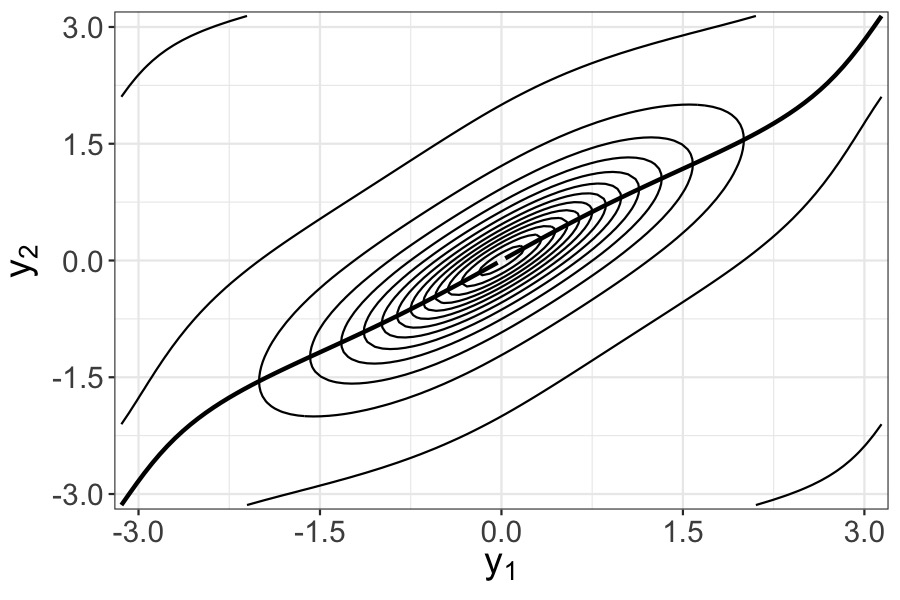}
    \caption{$\kappa_1 = \kappa_2 = 0.3$, $\rho = 0.4$.}
    \label{fig:circ2}
    \end{subfigure}
     \begin{subfigure}[b]{.48\textwidth}
    \centering
    \includegraphics[width=.8\textwidth]{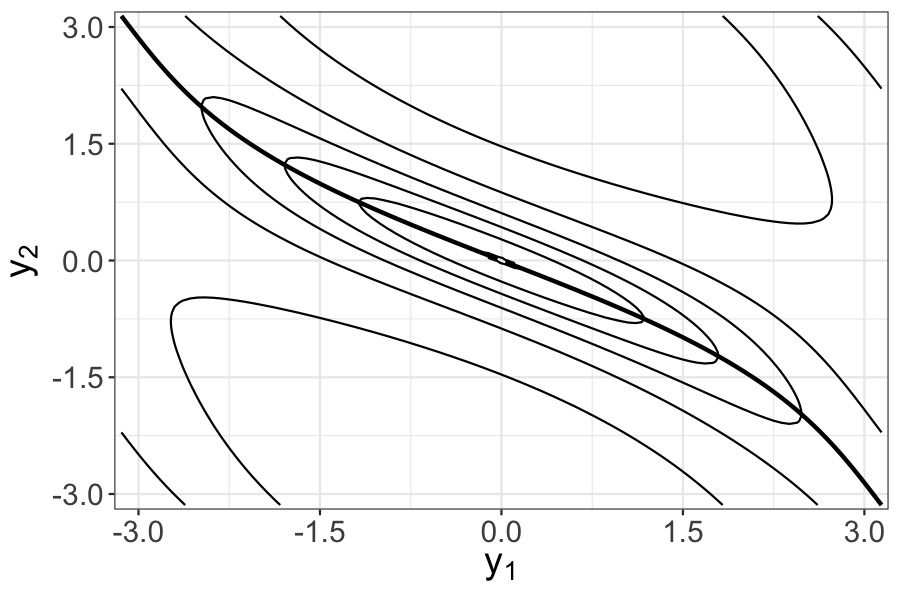}
    \caption{$\kappa_1 = 0.01$, $\kappa_2 = 0.2$, $\rho = -0.4$.}
    \label{fig:circ3}
    \end{subfigure}
    \begin{subfigure}[b]{.48\textwidth}
    \centering
    \includegraphics[width=.8\textwidth]{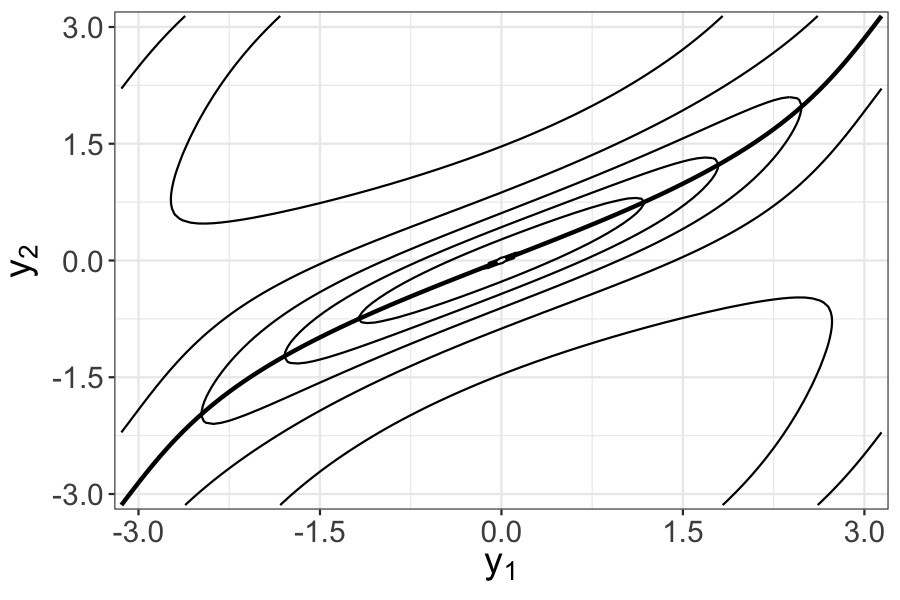}
    \caption{$\kappa_1 = 0.01$, $\kappa_2 = 0.2$, $\rho = 0.4$.}
    \label{fig:circ4}
    \end{subfigure}
    \caption{Contour plots of bivariate Wrapped Cauchy densities, centered at $\mu_1=\mu_2=0$ and obtained by varying the concentrations $\kappa_1,\kappa_2$ and the correlation parameter $\rho$. Contours are    projected on the plane obtained by unwrapping a torus. Curves (in bold) indicate the circular regression functions $\mathbb{E}(Y_2 \mid y_1)$. }
    \label{fig:wrappedCauchy}
\end{figure}
As mentioned in Section \ref{intro}, the literature offers a variety of toroidal densities that can be integrated in our model. The choice of a suitable parametric family should be a compromise between flexibility, ease of interpretation and numerical tractability. We rely on the particularly attractive bivariate wrapped Cauchy density  \citep{kato2015_toroidal} that optimally satisfies these three requirements. This density depends on a vector of five parameters, $\bm{\theta}=(\mu_1, \mu_2, \kappa_1, \kappa_2, \rho)$, namely two means, $-\pi \leq \mu_1,\mu_2\leq \pi$, two concentrations $0 \leq \kappa_1,\kappa_2 <1$ and a correlation parameter $-1 < \rho < 1$ and it can be written in terms of six coefficients $c, c_0 \ldots c_4$ that only depend on the three parameters $ \kappa_1,\kappa_2, \rho$ \citep{kato2015_toroidal}, namely
\begin{align} \label{eq:biv_cauchy}
f(\bm{y};\bm{\theta})=&\frac{c}{c_0-c_1 \cos(y_1-\mu_1)-c_2 \cos(y_2-\mu_2) -c_3 \cos(y_1-\mu_1)\cos(y_2-\mu_2)-c_4\sin(y_1-\mu_1)\sin(y_2-\mu_2)}.
\end{align}

The density shares several properties with the bivariate normal density: it is unimodal and pointwise symmetric about $(\mu_1, \mu_2)$; $\kappa_1$ ($\kappa_2$) controls the dispersion of  the marginal distribution of $y_1$ ($y_2$), which tends to a point mass at $\mu_1$ ($\mu_2$) when $\kappa_1$ ($\kappa_2$) approaches 1 and to a uniform distribution on the circle when $\kappa_1$ ($\kappa_2$) approaches 0; positive (negative) values of $\rho$  correspond to positive (negative) correlation between the two angles, with the density for $-\rho$ being the reflection of the density associated with $\rho$ and with $\rho=0$ corresponding to independence.

It furthermore depends on a normalizing constant that is available in closed form (a property not necessarily shared by other toroidal densities) and it is also closed under marginalization and conditioning (the two marginal and conditional densities are univariate wrapped Cauchy). This property not only facilitates the evaluation of regression lines of one circular variable over the other one, but it also makes simulation straightforward. A random toroidal observation can be obtained by drawing the first coordinate from the marginal distribution and then using this value for sampling the second coordinate from the univariate conditional distribution given the first coordinate. Since both the marginal and the conditional distributions are univariate wrapped Cauchy, the whole task can be undertaken by calls to a single simulation wrapper of this distribution such as the function {\tt rwrappedcauchy} in the R package {\tt circular} \citep{lund2017package}. Figure \ref{fig:wrappedCauchy} shows the shape taken by this density under specific parameter values and the relating regression functions.

\subsection{The latent process}
\begin{figure}
    \centering
    \includegraphics[scale=0.4]{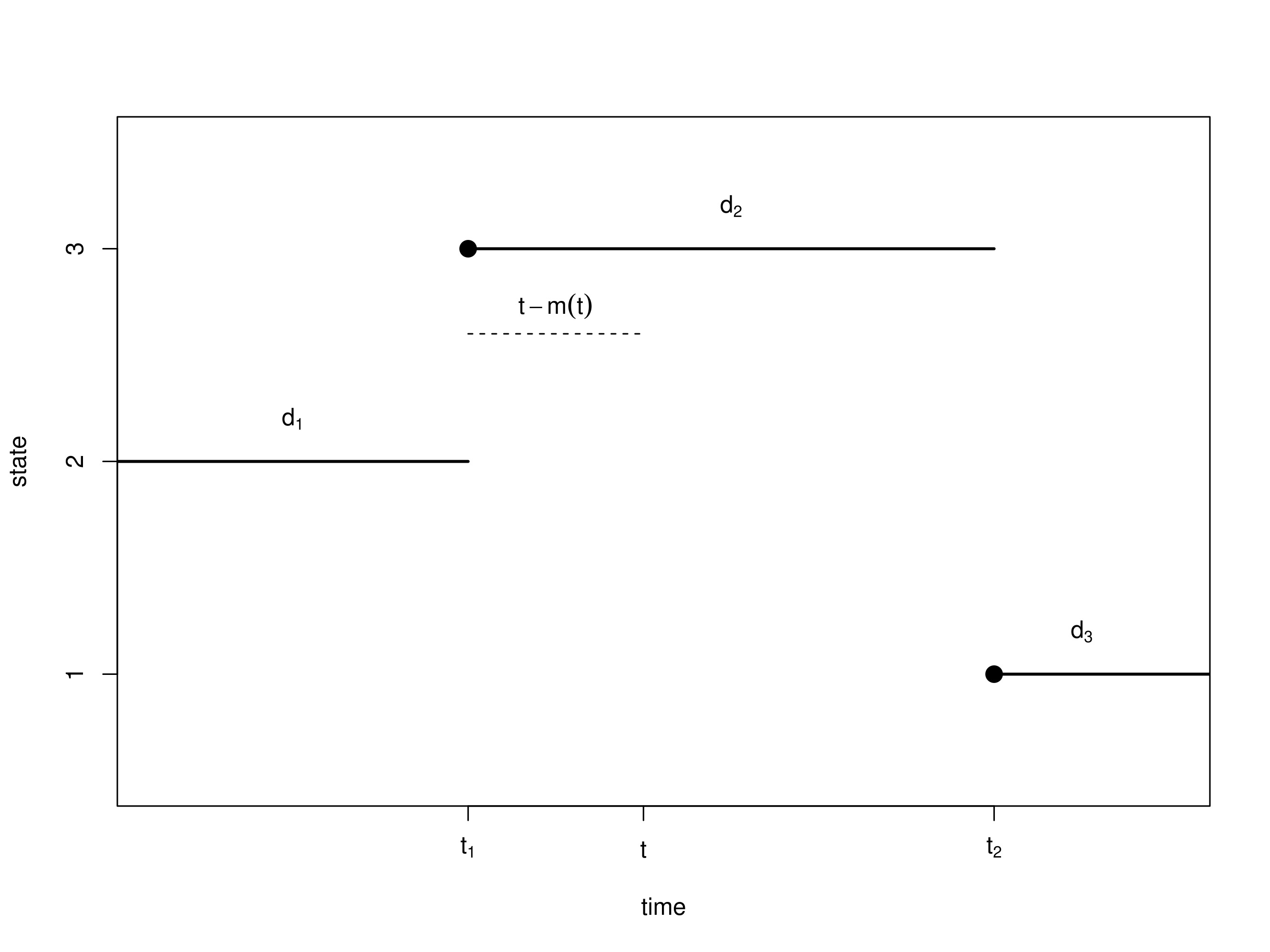}
    \caption{A sample path of a semi-Markov chain that changes state at random times $t_1$ and $t_2$, by remaining in the same state for dwell times $d_1$ and $d_2$. For each time $t$, $m(t)=\max\{t_n \mid t_n \leq t\}$ indicates the time of the last state change and $t-m(t)$ the time since the last state change.}
    \label{fig:hsmm_figure}
\end{figure}

The multinomial process $\bm{u}=(\bm{u}_t, t \geq 1)$  switches across states $1 \ldots K$ at random time points $t_n, n\in \mathbb{N}$, by remaining in the same state $\bm{u}_{t_n}$ during the dwell time $d_{n+1}=t_{n+1}-t_{n}$ (Figure \ref{fig:hsmm_figure} displays a typical sample path of the process). We assume that the evolution of the process is determined by the state of the process at the last label switch, namely
\begin{equation}p(\bm{u}_{t_{n+1}}, d_{n+1} \mid \bm{u}_{t_1} \ldots \bm{u}_{t_n}, d_1 \ldots d_n )=p(\bm{u}_{t_{n+1}}, d_{n+1}\mid \bm{u}_{t_n}).\label{eq:semi-markov hypothesis}
\end{equation}

\noindent Under \eqref{eq:semi-markov hypothesis}, the process is referred to as a semi-Markov chain \cite[equation (1.1)]{Barbu2008} and it is fully specified by the bivariate conditional distributions
\[p(\bm{u}_{t_{n+1}}, d_{n+1}\mid \bm{u}_{t_n})=p(d_{n+1} \mid \bm{u}_{t_{n+1}}, \bm{u}_{t_n})p(\bm{u}_{t_{n+1}} \mid \bm{u}_{t_n})=p(d_{n+1} \mid \bm{u}_{t_n})p(\bm{u}_{t_{n+1}}\mid \bm{u}_{t_n})\]
which can in turn be written as the product of a state-specific dwell time distribution
\begin{equation}\label{eq:dwell_time_distribution}p(d_{n+1} \mid \bm{u}_{t_n})=\prod_{k=1}^{K}p_k(d)^{u_{t_nk}}, \quad d=1,2, \ldots\end{equation} and a transition probability distribution $$p(\bm{u}_{t_{n+1}}\mid \bm{u}_{t_n})=\prod_{k=1}^K\prod_{h\neq k}\omega_{kh}^{u_{t_nk}u_{t_{n+1}h}},$$ where
$\omega_{kh}=P(u_{t_{n+1}h}=1 \mid u_{t_nk}=1)$
is the transition probability from state $h$ to state $k$, stored in a transition probability matrix $\bm{\Omega} = [\omega_{kh}]$, and the product of binary indicators
$u_{t_nk}u_{t_{n+1}h}$ indicates the transition event.

In a semi-Markov chain setting, dwell times are discrete and therefore the distributions $p_k(d)$ that appear in \eqref{eq:dwell_time_distribution}  can be  conveniently modelled by using life table tools. Specifically, or each time $t$, let $m(t)=\max\{t_n \mid t_n \leq t\}$ be the time of the last state change before $t$ (Figure \ref{fig:hsmm_figure}) and let the survival function
\[S_k(t-m(t))=P(D > t -m(t) \mid u_{m(t)k}=1)\]
be the probability of a dwell time $D$ larger than $t-m(t)$. Accordingly,
\begin{align}\label{eq:hazard.function}
q_k(t-m(t))=&P(t-m(t)< D\leq t-m(t)+1\mid D > t-m(t)) \nonumber\\
=&\frac{S_k(t-m(t))-S_k(t-m(t)+1)}{S_k(t-m(t))}=1-\frac{S_k(t-m(t)+1)}{S_k(t-m(t))}\end{align}
is the dwell time hazard function, that is the conditional probability of a state change at $t$ given that the chain is dwelling in state $k$ since time $m(t)$. Hazards \eqref{eq:hazard.function} are useful to obtain the survival function as  \begin{equation}\label{eq:survfromq}S_k(t-m(t))=\prod_{d=1}^{t-m(t)} (1-q_k(d)),\end{equation}
and the dwell time distribution, as
\begin{equation}\label{eq:densfromq}p_k(t-m(t))=q_k(t-m(t))S_k(t-m(t))=q_k(t-m(t))\prod_{d=1}^{t-m(t)} (1-q_k(d)).\end{equation}
Under this setting, if the chain is in state $k$ at time $t-1$, it will either remain in the same state with probability $1-q_k(t-m(t))$ or switch to a state $h \neq k$
with probability $q_k(t-m(t))\omega_{kh}$. Formally, the conditional distribution of the chain at time $t$, given the history of the process up to time $t-1$, is given by
\begin{equation}
p(\bm{u}_t \mid \bm{u}_s, s <t) =
   \prod_{k=1}^{K} \prod_{h \neq k}^{1 \ldots K}\left[1-q_k(t-m(t)\right]^{u_{tkh}}
    \left[q_k(t-m(t))\omega_{kh}\right]^{u_{tkh}},
\end{equation}
where $u_{tkh}=u_{t-1,k}u_{th}$ indicate the transition event from state $k$ to state $h$ at time $t$ and the joint distribution of the process is given by
\begin{align}p(\bm{u})\, = \, p(\bm{u}_1)\prod_{t=2}^{T}p(\bm{u}_t \mid \bm{u}_s, s < t) 
\, = \,
p(\bm{u}_1)\prod_{t=2}^{T}\prod_{k=1}^{K} \prod_{h \neq k}^{1 \ldots K}\left[1-q_k(t-m(t)\right]^{u_{tkh}}
    \left[q_k(t-m(t))\omega_{kh}\right]^{u_{tkh}},
    \label{eq:joint.distr.sm}\end{align}
where $p(\bm{u}_1)$ is the multinomial distribution of the initial state of the process.

In summary, the joint distribution of the semi-Markov chain depends on
$K$ mass probabilities of the initial distribution $p(\bm{u}_1)$, say $\pi_1 \ldots \pi_K$, $K(K-1)$ transition probabilities $\omega_{hk}$ and, finally, $K$ hazard functions $q_k(t)$. Because the hazard functions have potentially an infinite support, a popular approach relies on assuming that these hazards are parametric functions, say $q_k(t)=q_k(t,\bm{\beta})$, known up to a parametric vector $\bm{\beta}$ to be estimated.
Under this setting, the joint distribution of the latent process is fully parametric, say $p(\bm{u})=p(\bm{u}; \bm{\eta})$, and it is assumed known up to the parametric vector $\bm{\eta}=(\bm{\pi},\bm{\omega},\bm{\beta})$.

\subsection{A proportional hazards model for the dwell time}
 A parametric model of the dwell time distribution can be equivalently specified by choosing either a parametric family of probability distributions $p_k(t; \bm{\beta})$ or a family of  hazard functions $q_k(t; \bm{\beta})$, because of the one-to-one correspondence between distributions of positive random variables and their hazard functions. In the HSMM literature, the first approach is often preferred and relies on the choice of discrete distributions such as the (shifted) Poisson or negative binomial distribution \citep{Bulla2006,Economou2014,vandeKerk2015}.  The second approach is instead preferred in the survival analysis literature. The choice between the two approaches depends on whether the focus is either on the conditions under which the dwell time distribution $p_k(t; \bm{\beta})$ is geometric or not, or on the conditions under which the chances of a regime switch, measured by  $q_k(t; \bm{\beta})$, are time-constant or time-varying.

 In marine studies, the interest is on understanding whether the risk associated with a shift in sea conditions is time-constant or time-varying and a survival analysis approach seems the most natural approach. Getting therefore away from the mainstream HSMM literature, we focus on hazard modelling and assume that
\[g(q_k(t-m(t)))=\beta_{0k}+\beta_{1k}(t-m(t)),\]
where $g(\cdot)$ is a suitable link function that maps the hazard to a linear function of time. The equation above can be extended by introducing a row profile of (possibly time-varying) covariates, say
\[g(q_k(t-m(t)))=\beta_{0k}+\beta_{1k}(t-m(t))+\bm{x}_t^{\sf T}\bm{\beta}_k.\]
When the regression parameters $\beta_{1k}$ and $\bm{\beta}_k$
are equal to zero, the hazard is time-constant and the dwell time distribution is geometric. Otherwise, these regression parameters capture the effect of time and other time-varying covariates in the link scale.

 We exploit a complementary log-log transformation, leading to the model
\begin{equation}\label{eq:cloglog}\log(-\log(1-q_k(t-m(t)\mid \bm{x}^{\sf T})))\approx \beta_{0k}+\beta_{1k}(t-m(t)-0.5)+\bm{x}_t^{\sf T}\bm{\beta}_k,\end{equation}
which is the discrete-time counterpart of a continuous-time proportional hazards model \citep[pag. 37]{Kalbfleisch1980}. It is obtained by discretizing a continuous hazard function
\begin{equation}\label{eq:continuous.hazard}h_k(t-m(t); \bm{x}^{\sf T}_t)=h_{k0}(t)\exp (\bm{x}_t^{\sf T}\bm{\beta}_k)= \exp(\beta_{0k}+\beta_{1k}(t-m(t))+\bm{x}_t^{\sf T}\bm{\beta}_k) \quad t \in (0, +\infty),\end{equation}
specified by the product of a baseline hazard $h_{k0}(t)$ and an exponential function of time-varying covariates, fullfilling a traditional proportional hazard assumption. Under
\eqref{eq:continuous.hazard}, the survival function is given by \[S_k(t-m(t)\mid \bm{x}_t^{\sf T})=\exp\left(-\int_0^{t-m(t)}h_k(\tau -m(t)) \exp\left(\bm{x}_\tau^{\sf T}\bm{\beta}\right)\right)d \tau,\]
and
\[q_k(t-m(t))=1-\exp \int_{t-m(t)-1}^{t-m(t)}h_k(\tau-m(t)\mid \bm{x}_t^{\sf T})d \tau \approx 1- \exp(h_k(\tau-m(t)-0.5\mid \bm{x}_t^{\sf T})),\]
leading to \eqref{eq:cloglog}.

\section{Data augmentation and maximum likelihood estimation}\label{sec:estimation}
Maximization of the likelihood function of the proposed hierarchical model \eqref{eq:hierachical_model} is complicated by the summation over all the possible paths of the latent semi-Markov chain. This is a typical difficulty in the literature of hierarchical models and it is often overcome by an EM algorithm that alternates the maximization of a weighted complete-data log-likelihood function (M step) with weights updates (E step). The complete-data log-likelihood function is defined by appropriately augmenting the latent variables by pseudo-observations. The efficiency of the EM algorithm depends on the augmentation design, because only an appropriate definition of the pseudo-observations facilitates the execution of both the E step and the M step of the algorithm.

We use an augmentation that is obtained by representing a semi-Markov chain as an augmented Markov chain. To illustrate, we observe that when hazards do not depend on the dwell time, say $q_k(t)=q_k$, the survival function of the dwell time is geometric, $S_k(t-m(t))=(1-q_k)^{t-m(t)}$ and the joint distribution \eqref{eq:joint.distr.sm} reduces to the distribution of a Markov chain with transition probabilities
\[\pi_{kh}=P(u_{th}=1 \mid u_{t-1,k}=1)=\begin{cases}
    1-q_k & h=k \\
    q_k\omega_{kh} & h \neq k
\end{cases},\]
namely
\begin{equation}
p(\mathbf{u})=p(\bm{u}_1)\prod_{t=2}^{T}\prod_{k=1}^{K} \prod_{h =1}^{K}\pi_{kh}^{u_{tkh}}. \label{eq:mc_distr}
\end{equation}
A compact form of the joint distribution \eqref{eq:joint.distr.sm}, similar to the equation \eqref{eq:mc_distr} above, can be also derived when hazards are not necessarily time-constant, by suitably augmenting the observed path of the process \citep{Anselone1960}. Precisely, let $M=\max\{d_n, n \in \mathbb{N}\}$ be the maximum  dwell time of the semi-Markov chain. We first augment the multinomial vector $\bm{u}_t$ that represents the state of the chain at time $t$ by a multinomial vector that includes the dwell time information, say $(u_{tkhd},k,h =1 \ldots K, d=1\ldots M)$, where $u_{tkhd}=1$ if the chain moves to state $h$ after having dwelled in state $k$ for a period of $d$ times, and $0$ otherwise. Second, we arrange the $K\times K \times M$ vector elements as entries of a block matrix of $K \times K$ square blocks of dimension $M \times M$, say  $\bm{U}_t$, with diagonal blocks
\[\bm{U}_{tkk}=\left(\begin{array}{ccccc}
0 & u_{tkk2} & 0 & \ldots & 0\\
0 & 0 & u_{tkk3} & \ldots & 0\\
\ldots & \ldots & \ldots & \ldots & \ldots\\
0 & 0 & 0 & \ldots & u_{tkk(M-1)}\\
0 & 0 & 0 & \ldots & u_{tkkM}
\end{array}\right)\]
and off-diagonal blocks
\[\bm{U}_{tkh}=\left(\begin{array}{ccccc}
u_{tkh1} & 0& 0 & \ldots & 0\\
u_{tkh2}& 0& 0 & \ldots & 0\\
\ldots & \ldots & \ldots & \ldots & \ldots\\
u_{tkh(M-1)}& 0& 0 & \ldots & 0\\
u_{tkhM}& 0& 0 & \ldots & 0\\
\end{array}\right).\]
Third, for each $t >1$ we introduce a block matrix of $K \times K$ square blocks of dimension $M \times M$, say
\[\bm{\Gamma}_t = \left(\bm{\Gamma}_{tkh}, k,h=1 \ldots K\right),\]
where the diagonal blocks are given by
\begin{equation}\label{eq:gammadiag}
\Gamma_{tkk}
=\left(\begin{array}{ccccc}
0 & \gamma_{tkk1} & 0 & \ldots & 0\\
0 & 0 & \gamma_{tkk2} & \ldots & 0\\
\ldots & \ldots & \ldots & \ldots & \ldots\\
0 & 0 & 0 & \ldots & \gamma_{tkk,M-1}\\
0 & 0 & 0 & \ldots & \gamma_{tkkM}
\end{array}\right)
\end{equation}
with $\gamma_{tkkd}=(1-q_k(d; \bm{x}^{\sf T}_t))$ and the off-diagonal blocks are given by
\begin{equation}\label{eq:gammanondiag}
    \Gamma_{tkh}=\left(\begin{array}{ccccc}
\gamma_{tkh1} & 0& 0 & \ldots & 0\\
\gamma_{tkh2}& 0& 0 & \ldots & 0\\
\ldots & \ldots & \ldots & \ldots & \ldots\\
\gamma_{tkh,M-1}& 0& 0 & \ldots & 0\\
\gamma_{tkhM}& 0& 0 & \ldots & 0\\
\end{array}\right),
\end{equation}
with $\gamma_{tkhd}=\omega_{kh}q_k(d ; \bm{x}_{t}^{\sf T})$.
Under this setting, the joint distribution of the semi-Markov chain can be written as
\begin{equation}
\label{eq:semi-markov.chain.compactdistr}
p(\mathbf{u})=p(\bm{u}_1)\prod_{t=2}^{T}\prod_{h=1}^{K}\prod_{k=1}^{K} \prod_{d=1}^M\gamma_{tkhd}^{u_{tkhd}}.
\end{equation}

We note that the probabilities $\gamma_{tkhd}$ are explicit functions of the hazards $q_k$, which is turn depend on the dwell time distributions $p_k$, via \eqref{eq:densfromq}. This supports the choice of directly modelling the hazards instead of modelling the dwell time distribution, as it is normally done in the HSMM literature.

We also note that \eqref{eq:semi-markov.chain.compactdistr} requires the knowledge of the maximum dwell time $M$, an information that we do not have when the process is latent. Remarkably, however, \eqref{eq:semi-markov.chain.compactdistr} is an approximation of the true joint distribution when $M$ is less than the maximum dwell time \citep{Langrock2011}. Accordingly, the function obtained by integrating \eqref{eq:conditional_data_distribution} and \eqref{eq:semi-markov.chain.compactdistr}, namely
\begin{equation}\label{eq:M_likelihood}
L(\bm{\theta},\bm{\eta})=L(\bm{\theta},\bm{\pi},\bm{\omega},\bm{\beta})=
\sum_{\bm{u}}\prod_{t=1}^{T}\prod_{k=1}^{K}f(\bm{y}_t; \bm{\theta}_k)^{u_{tk}}p(\bm{u}_1;\bm{\pi})\prod_{t=2}^{T}\prod_{k=1}^{K}\prod_{h=1}^{H}\prod_{d=1}^{M}\gamma_{tkhd}(\bm{\omega},\bm{\beta})^{u_{tkhd}}
\end{equation}
is either the exact likelihood function (if $M$ is not less than the maximum dwell time), or it is otherwise an approximation of the exact likelihood, whose accuracy can be however indefinitely improved by increasing $M$.

The likelihood function \eqref{eq:M_likelihood}  can be maximized by an EM algorithm that relies on  the following complete-data log-likelihood function
\begin{eqnarray}\label{eq:complete.loglikelihood}\log L_{\textrm{comp}}(\bm{\theta},\bm{\pi},\bm{\omega},\bm{\beta})&=&\sum_{k=1}^{K}u_{1k}\log \pi_k+\sum_{t=2}^{T}\sum_{k=1}^{K}\sum_{h=1}^{K}\sum_{d=1}^{M}u_{tkhd}\log \gamma_{thkd}(\bm{\omega},\bm{\beta}) 
+\sum_{t=1}^{T}\sum_{k=1}^{K}u_{tk}\log f(\bm{y}_t;\bm{\theta}_{k}) .\end{eqnarray}
The algorithm is iterated by alternating an expectation (E) and a maximization (M) step. Given the parameter estimates, obtained at the end of the $s$-th iteration, the $(s+1)$-th iteration is initialized by the E-step, which evaluates the expected value of the complete data log-likelihood (\ref{eq:complete.loglikelihood}) with respect to the conditional distribution of the missing values $u_{tk}$ and $u_{tkhd}$ given the observed data. The E step reduces to the computation of the univariate posterior probabilities
 $\hat{\pi}_{tk} =P(u_{tk}=1\mid \bm{y})$, $t=1 \ldots T, k=1 \ldots K$ and the trivariate posterior probabilities $\hat{\pi}_{tkhd}=P(u_{t-1,k}=1, u_{th}=1, t-m(t)=d  \mid  \bm{y})$,  $t=2 \ldots T$, $k,h=1 \ldots K$, $d=2 \ldots M$.
The task of computing these posterior probabilities is generally referred to as the smoothing numerical issue and it is typically solved by   specifying the posterior probabilities in terms of suitably normalized functions, which can be computed recursively, avoiding unpractical summations over the state space of latent semi-Markov chain and numerical under- and over-flows. We used the smoothing algorithm discussed by
\citet{Cappe_etal2005}, in Section 5.1.1.

The M-step of the algorithm updates the parameter estimates, by maximizing the expected value of the complete data log-likelihood (\ref{eq:complete.loglikelihood}), obtained from the previous E step. This expected value is the sum of functions that depend on independent sets of parameters and can therefore be maximized separately. Specifically, the expected log-likelihood function
\begin{eqnarray}\label{eq:exp.complete.loglikelihood}Q(\bm{\theta}, \bm{\pi},\bm{\omega},\bm{\beta})&=&Q(\bm{\pi},\bm{\omega},\bm{\beta})+Q(\bm{\theta}) \nonumber\\
&=&\sum_{k=1}^{K}\hat{\pi}_{1k}\log \pi_k+\sum_{t=2}^{T}\sum_{k=1}^{K}\sum_{h=1}^{K}\sum_{d=2}^{M} \hat{\pi}_{tkhd}\log \gamma_{tkhd}(\bm{\omega},\bm{\beta}) 
\,+\,\sum_{t=1}^{T}\sum_{k=1}^{K}\hat{\pi}_{tk}\log f(\bm{y}_t;\bm{\theta}_{k}).\end{eqnarray}
\noindent The function
\[Q(\bm{\pi},\bm{\omega},\bm{\beta})=\sum_{k=1}^{K}\hat{\pi}_{1k}\log \pi_k+\sum_{t=2}^{T}\sum_{k=1}^{K}\sum_{h=1}^{K}\sum_{d=2}^{M}\hat{\pi}_{tkhd}\log \gamma_{tkhd}(\bm{\omega},\bm{\beta})\] can be split as the sum of three components that depend on independent parameters and that can be then maximized separately:
\begin{equation}\label{eq:q.initial}
Q(\bm{\pi})=\sum_{k=1}^{K}\hat{\pi}_{1k}\log \pi_k,
\end{equation}
\begin{equation}\label{eq:Q.omega}
Q(\bm{\omega})=\sum_{t=2}^{T}\sum_{k=1}^{K}\sum_{h \neq k}\sum_{d=2}^{M} \hat{\pi}_{tkhd}\log \omega_{kh}\end{equation}
and
\begin{equation}
Q(\bm{\beta})=\sum_{t=2}^{T}\sum_{k=1}^{K}\sum_{h \neq k}\sum_{d=2}^{M} \hat{\pi}_{tkhd}\log q_k(d;\bm{x}^{\sf T}_t,\bm{\beta}) +\sum_{t=2}^{T}\sum_{k=1}^{K}\sum_{d=2}^{M} \hat{\pi}_{tkkd}\log (1-q_k(d; \bm{x}^{\sf T}_t,\bm{\beta}))\label{eq:Q.hazard}
\end{equation}
Functions $Q(\bm{\pi})$ and $Q(\bm{\omega})$ are weighted multinomial loglikelihoods and are respectively maximized by $\hat{\pi}_k=\hat{\pi}_{1k}$
and
\[\hat{\omega}_{kh}=\frac{\sum_{t=2}^{T}\sum_{d=1}^{M} \hat{\pi}_{tkhd}}{\sum_{k \neq h}\sum_{t=2}^{T}\sum_{d=2}^{M} \hat{\pi}_{thkd}}.\]
The function $Q(\bm{\beta})$ is instead a weighted binomial log-likelihood and, accordingly, can  be  maximized by fitting a binomial regression model with weights and an appropriate link function. The choice of the transformation depends on the model choice for the hazard  function. Under the  proportional hazards assumption formulated in Section \ref{sec:model}, $Q(\bm{\beta})$ is maximized by a weighted binomial regression model that uses the complementary log-log link.
Finally, maximization of the function $Q(\bm{\theta})$ can be undertaken by an unconstrained maximization algorithm, after a suitable re-parametrization of the involved parameters. In the case of the bivariate wrapped Cauchy, for example, we
maximize $Q(\bm{\theta})$ over the parameter vector $(\theta_1, \theta_2, \theta_3, \theta_4, \theta_5)$ where $\theta_1=\tan(\mu_1)$,  $\theta_2=\tan(\mu_2)$, $\theta_3=(\textrm{tanh}(\kappa_1)+1)/2$, $\theta_4=(\textrm{tanh}(\kappa_2)+1)/2$ and, finally, $\theta_5=\textrm{tanh}(\rho)$, exploiting a quasi-Newton procedure as that provided, for example, by the function \texttt{optim} in R.

\section{Computational details, uncertainty quantification and model selection}
\label{sec:comp.details}
It is well known that the EM algorithm may converge to local maxima of the log-likelihood function or singularities at the edge of the parameter space, where the log-likelihood is unbounded. As a result, several strategies have been proposed to select a local maximizer and detect a spurious maximizer
\citep{maruotti2021_em}. We pursue a short-run strategy, by running the EM algorithm from a number of random initializations, and stop it without waiting for full convergence.
Then, we select the best output across these short runs, i.e. the one maximizing the log-likelihood, and use this solution to initialize longer runs to reach convergence. Eventually, we stop the optimization when the increase of two successive log-likelihoods fell below $10^{-6}\%$, as this stopping criterion produced stable parameter estimates in preliminary experiments. The computational burden of a single EM run increases with $K$ and $M$, which directly determine the size of the matrices in \eqref{eq:gammadiag} and \eqref{eq:gammanondiag}. Although the proposed algorithm does not involve matrix inversions or determinant computations, large matrices may require the allocation of a large portion of storage memory.

The EM algorithm does not provide information about the sampling distribution of the parameters estimates and approximations based on the observed information matrix often require a very large sample size. As a result, we rely on a bootstrap strategy, which is  convenient in this context because the simulation of the proposed HSMM is straightforward. Precisely, we run the EM algorithm for $B = 1000$ times, obtaining $\lbrace \bm{\eta} \rbrace_{b=1}^B$ and $\lbrace \bm{\theta} \rbrace_{b=1}^B$ sets of estimates, which we exploited to compute standard errors.

Model selection reduces to find the optimal number $K$ of components, according to a suitable criterion. Like any other mixture model, our proposal can be either exploited within a density estimation context, where the goal is a good approximation of the data density, or under a clustering framework, where the goal is good data segmentation \citep{FruhwirthSchnatter2018}. In our case study, segmentation is more important than density estimation and we therefore rely on the integrated complete likelihood \citep[ICL;][]{Biernacki2000} where goodness of fit is penalized by clusters overlaps.


\section{Simulation  study}\label{sec:simulation}
\begin{figure}[ht]
     \centering
     \begin{subfigure}[b]{.32\textwidth}
     \centering
     \includegraphics[width=0.95\textwidth]{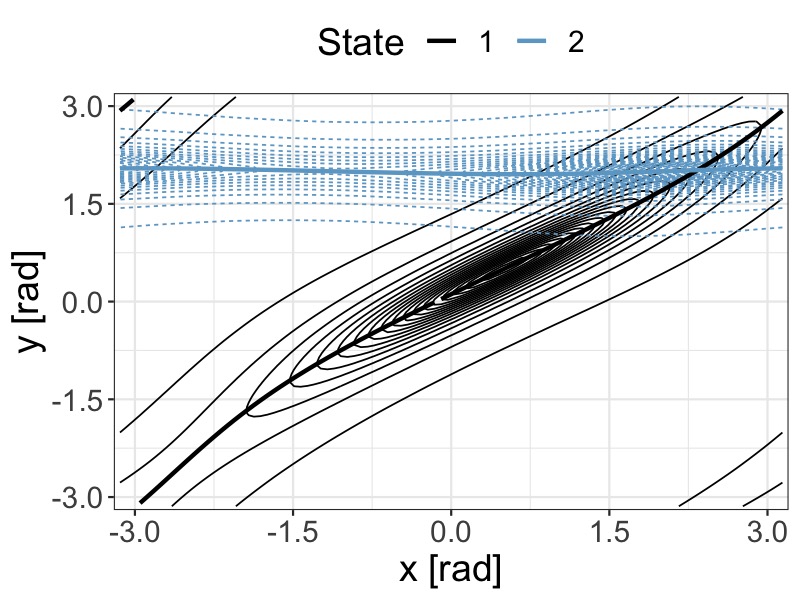}
   \caption{$K = 2$}
 		\label{fig:simdens2}
     \end{subfigure}
      \begin{subfigure}[b]{.32\textwidth}
     \centering
     \includegraphics[width=0.95\textwidth]{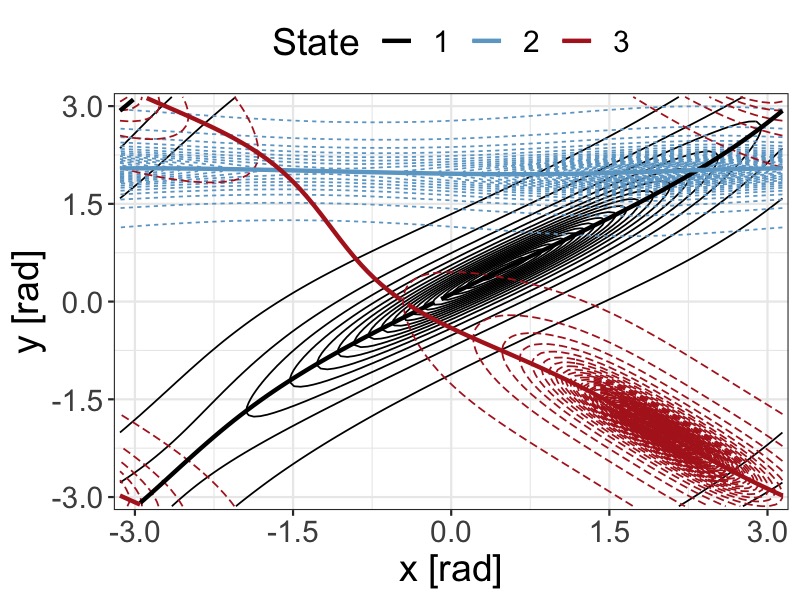}
   \caption{$K = 3$}
 		\label{fig:simdens3}
     \end{subfigure}
      \begin{subfigure}[b]{.32\textwidth}
     \centering
     \includegraphics[width=0.95\textwidth]{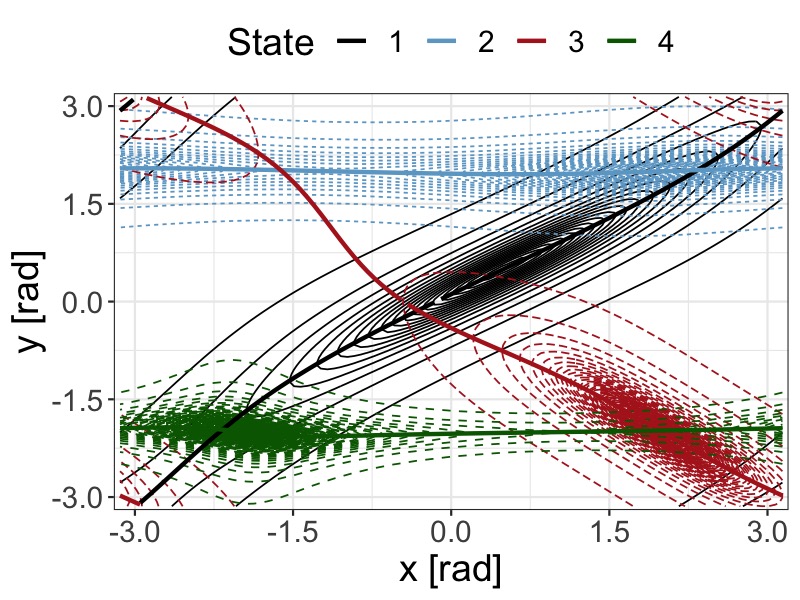}
   \caption{$K = 4$}
 		\label{fig:simdens4}
     \end{subfigure}
     \caption{The three scenarios considered for the simulation study}
     \label{fig:simdens}
 \end{figure}

We run a simulation experiment to assess the ability of our proposal to recover the population parameter values under different scenarios, as well as to assign each observation a probability of coming from one of the latent regimes, and eventually classify it. The simulation was performed under $K \in \lbrace 2, 3, 4\rbrace$ regimes and included a time-varying covariate, drawn from $N(0, 9)$. Table \ref{tab:truepars} and Figure \ref{fig:simdens} respectively display the population values of the parameters chosen for simulating three scenarios, and the contour plots of the resulting component densities. These scenarios -- though not exhaustive -- aim to mimic possible real data situations where mixture components may (or may not) overlap and show positive (or negative) correlations, while also accounting for different effects of a time-varying covariate on the dwell-time distribution and not necessarily uniform transition probabilities.

\begin{figure}[ht]
    \centering
    \begin{subfigure}[b]{.32\textwidth}
    \centering
    \includegraphics[width=.95\textwidth]{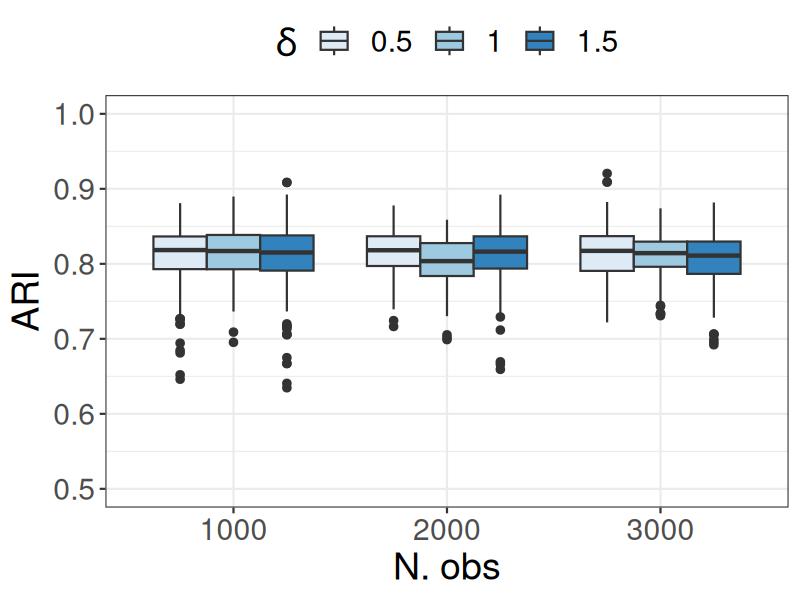}
    \caption{$K = 2$}
    \end{subfigure}
     \begin{subfigure}[b]{.32\textwidth}
    \centering
    \includegraphics[width=.95\textwidth]{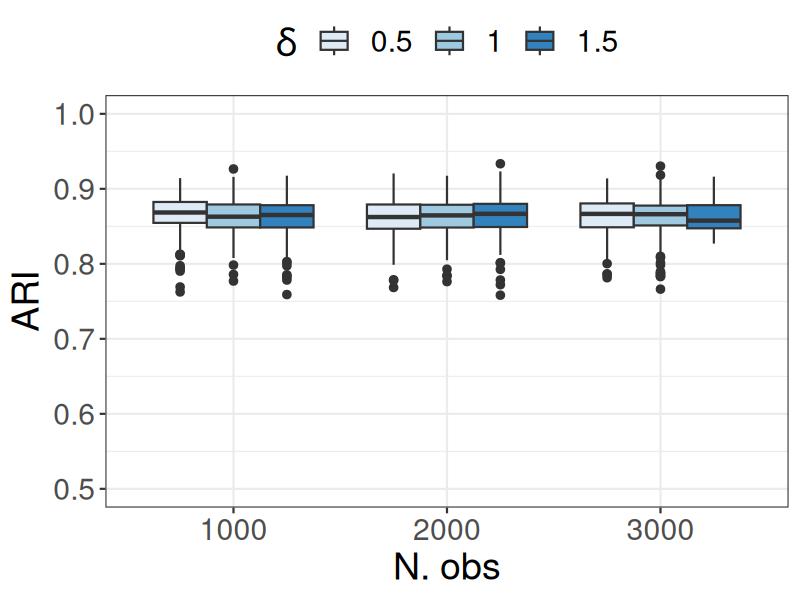}
    \caption{$K = 3$}
    \end{subfigure}
     \begin{subfigure}[b]{.32\textwidth}
    \centering
    \includegraphics[width=.95\textwidth]{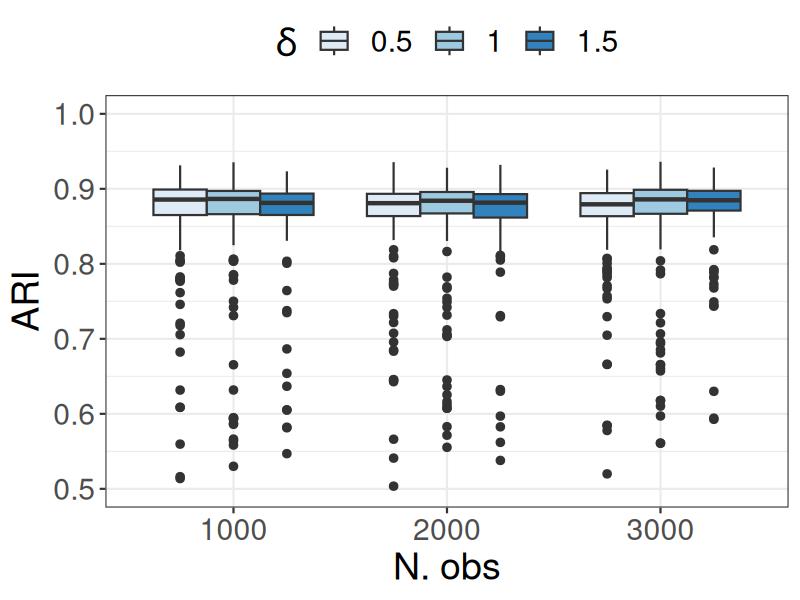}
    \caption{$K = 4$}
    \end{subfigure}
    \caption{Simulation study: classification performances in terms of ARI in each considered scenario.}
    \label{fig:ARI}
\end{figure}

For each $K$, we simulated $N = 250$ samples with increasing size  $T \in \lbrace 1000, 2000, 3000\rbrace$ to evaluate the consistency of the estimates. For each simulated sample, we know the maximum dwell time $M_{\text{obs}}$ and such knowledge provides us with the opportunity to perform a sensitivity analysis of the geometric approximation of the dwell time distribution. Accordingly, parameter estimation was repeated by maximizing the log-likelihood \eqref{eq:M_likelihood} with $M = \lceil M_{\text{obs}}\times \delta \rceil$, with $\delta \in \lbrace 0.5, 1, 1.5 \rbrace$. The performance of the proposed estimation algorithm was finally assessed in terms of (i) Adjusted Rand Index (ARI) for the classification of the latent regimes and (ii) Root Mean Squared Error (RMSE) for the estimation of the parameters of both the latent part of the model, $\bm{\eta}$, and the observed part of the model, $\bm{\theta}$.  Specifically, the RMSE associated with the circular means was computed by using the angular deviation \citep{mardia2000directional} of the estimates. The results are summarized by Figure \ref{fig:ARI} and Tables Tables \ref{tab:tab2}-\ref{tab:tab4}.

In this study, the median ARI was always above 0.8 (Figure \ref{fig:ARI}), reassuring  about the capability of the algorithm to perform a satisfactory data segmentation, even when the sample size is small and the maximum dwell time $M$ is misspecified. In particular, the rightmost panel of the figure shows that most ARI values are above 0.8 also when the model dimension is large and the algorithm is challenged by a huge number of parameters. Under this setting, however, we obtained a few cases of incorrect data segmentation, associated with a sub-optimal local maximum of the log-likelihood. It is possible that the chances of  suboptimal solutions increase with model complexity.

The RMSE values shown by Tables \ref{tab:tab2}-\ref{tab:tab4} alway decrease with sample size, indicating the consistency of the estimates obtained by the proposed algorithm. When the log-likelihood is misspecified by a value of $M$ that is smaller than the true maximum dwell time, the RMSE is always larger, as expected. This difference is attenuated when instead $M$ is overestimated, suggesting the somehow obvious strategy of choosing the largest $M$ allowed by the available storage memory.


\begin{table}[ht]
\setlength\extrarowheight{-2pt}
    \centering
		\resizebox{.8\textwidth}{!}{%
		 \begin{tabular}{cc|ccccc|ccc|cccc}
    \toprule
    && \multicolumn{5}{c}{observation process} & \multicolumn{7}{c}{latent process} \\
    && &&&&&\multicolumn{3}{c}{dwell time}& \multicolumn{4}{c}{transition probabilities}\\
    no. states & state $k$ &  $\mu_{1k}$ &  $\mu_{2k}$ &  $\kappa_{1k}$ &  $\kappa_{2k}$ &  $\rho_{k}$ & $\beta_{0k}$  &  $\beta_{1k}$ &  $\beta_{2k}$& $\omega_{k1}$ & $\omega_{k2}$ & $\omega_{k3}$ & $\omega_{k4}$\\
       \midrule
       \multirow{2}{*}{2} & 1 & 0.5 & 0.5 & 0.2 & 0.3 & 0.6 & -8 & 0.35 & -0.5&0 & 1 & -- & -- \\
       & 2  & 2 & 2 & 0.2 & 0.8 & 0.1 & -3 & 0.075 & 0.5& 1 & 0 & -- & --\\
       \midrule
       \multirow{3}{*}{3} & 1 & 0.5 & 0.5 & 0.2 & 0.3 & 0.6 &-8 & 0.4 & -0.5& 0 & 0.50 & 0.50& --\\
       & 2 & 2 & 2 & 0.2 & 0.8 & 0.1 &-5 & 0.15 & 0.2& 0.90 & 0 & 0.10 & -- \\
       & 3 & 2 & -2 & 0.5 & 0.5 & -0.6 & -3 & 0.05 & 0.7& 0.45 & 0.55 & 0& --\\
       \midrule
       \multirow{4}{*}{4} & 1 & 0.5 & 0.5 & 0.2 & 0.3 & 0.6 &  -8 & 0.4 & -0.5& 0 &0.25 & 0.25 & 0.50  \\
       &2 & 2 & 2 & 0.2 & 0.8 & 0.1 & -6 & 0.3 & 0.2& 0.70 & 0 & 0.20 & 0.10\\
       &3 & -2 & -2 & 0.7 & 0.9 & -0.3 & -4 & 0.05 & 0.7& 0.15 & 0.25 & 0 & 0.60\\
       &4 & 2 & -2 & 0.5 & 0.5 & -0.6 & -2 & 0.15 & -0.1& 0.30 & 0.20 & 0.50 & 0\\
       \bottomrule
    \end{tabular}
    }
    \caption{Simulation: parameter values chosen for each  scenario.}
    \label{tab:truepars}
\end{table}

\begin{table}[p]
\setlength\extrarowheight{-1pt}
\centering
\scriptsize
\begin{tabular}{lcccc|ccc|ccc}
\toprule%
& $K$ & \multicolumn{3}{c}{$2$} & \multicolumn{3}{c}{$3$} &
\multicolumn{3}{c}{$4$} \\
\cline{2-5}\cline{6-8}\cline{9-11}%
& $T$ & $\delta = 0.5$ & $\delta = 1$ & $\delta = 1.5$ & $\delta = 0.5$ & $\delta = 1$ & $\delta = 1.5$ & $\delta = 0.5$ & $\delta = 1$ & $\delta = 1.5$  \\
\midrule
\multirow{3}{*}{$\mu_{11}$} & 1000 & 0.087 & 0.085 & 0.083 & 0.110 & 0.121 & 0.107 & 0.149 & 0.118 & 0.140 \\
 & 2000 & 0.062 & 0.062 & 0.063 & 0.070 & 0.076 & 0.074 & 0.089 & 0.087 & 0.086 \\
 & 3000 & 0.054 & 0.059 & 0.057 & 0.076 & 0.070 & 0.063 & 0.081 & 0.070 & 0.078 \\
\midrule
\multirow{3}{*}{$\mu_{12}$} & 1000 & 0.146 & 0.162 & 0.148 & 0.166 & 0.189 & 0.189 & 0.311 & 0.205 & 0.253 \\
 & 2000 & 0.135 & 0.133 & 0.133 & 0.104 & 0.110 & 0.105 & 0.203 & 0.146 & 0.145 \\
 & 3000 & 0.098 & 0.098 & 0.099 & 0.089 & 0.083 & 0.088 & 0.159 & 0.116 & 0.161 \\
 \midrule
\multirow{3}{*}{$\mu_{13}$} & 1000 & -- & -- & -- & 0.097 & 0.097 & 0.107 & 0.397 & 0.028 & 0.383 \\
 & 2000 & -- & -- & -- & 0.068 & 0.063 & 0.068 & 0.314 & 0.022 & 0.247 \\
 & 3000 & -- & -- & -- & 0.055 & 0.055 & 0.056 & 0.309 & 0.018 & 0.246 \\
  \midrule
 \multirow{3}{*}{$\mu_{14}$} & 1000 & -- & -- & -- & -- & -- & -- & 0.442 & 0.087 & 0.398 \\
 & 2000 & -- & -- & -- & -- & -- & -- & 0.289 & 0.057 & 0.239 \\
 & 3000 & -- & -- & -- & -- & -- & -- & 0.308 & 0.046 & 0.235 \\
  \midrule
 \multirow{3}{*}{$\mu_{21}$} & 1000 & 0.074 & 0.073 & 0.071 & 0.094 & 0.101 & 0.093 & 0.149 & 0.103 & 0.124 \\
 & 2000 & 0.053 & 0.054 & 0.055 & 0.011 & 0.011 & 0.010 & 0.076 & 0.074 & 0.073 \\
 & 3000 & 0.047 & 0.052 & 0.049 & 0.009 & 0.009 & 0.009 & 0.070 & 0.059 & 0.068 \\
\midrule
\multirow{3}{*}{$\mu_{22}$} & 1000 & 0.019 & 0.019 & 0.019 & 0.176 & 0.013 & 0.171 & 0.223 & 0.019 & 0.130 \\
 & 2000 & 0.015 & 0.014 & 0.014 & 0.011 & 0.011 & 0.011 & 0.143 & 0.014 & 0.014 \\
 & 3000 & 0.011 & 0.011 & 0.011 & 0.010 & 0.011 & 0.010 & 0.116 & 0.012 & 0.116 \\
 \midrule
\multirow{3}{*}{$\mu_{23}$} & 1000 & -- & -- & -- & 0.200 & 0.097 & 0.209 & 0.366 & 0.008 & 0.270 \\
 & 2000 & -- & -- & -- & 0.067 & 0.063 & 0.067 & 0.173 & 0.006 & 0.054 \\
 & 3000 & -- & -- & -- & 0.058 & 0.055 & 0.058 & 0.208 & 0.005 & 0.130 \\
  \midrule
 \multirow{3}{*}{$\mu_{24}$} & 1000 & -- & -- & -- & -- & -- & -- & 0.097 & 0.091 & 0.094 \\
 & 2000 & -- & -- & -- & -- & -- & -- & 0.066 & 0.056 & 0.060 \\
 & 3000 & -- & -- & -- & -- & -- & -- & 0.063 & 0.046 & 0.052 \\
  \midrule
 \multirow{3}{*}{$\kappa_{11}$} & 1000 & 0.023 & 0.023 & 0.022 & 0.031 & 0.026 & 0.030 & 0.046 & 0.033 & 0.044 \\
 & 2000 & 0.018 & 0.018 & 0.018 & 0.021 & 0.020 & 0.021 & 0.028 & 0.024 & 0.029 \\
 & 3000 & 0.015 & 0.014 & 0.014 & 0.017 & 0.016 & 0.016 & 0.025 & 0.020 & 0.023 \\
\midrule
\multirow{3}{*}{$\kappa_{12}$} & 1000 & 0.045 & 0.045 & 0.044 & 0.051 & 0.033 & 0.049 & 0.063 & 0.048 & 0.050 \\
 & 2000 & 0.029 & 0.028 & 0.028 & 0.024 & 0.023 & 0.024 & 0.045 & 0.030 & 0.031 \\
 & 3000 & 0.025 & 0.024 & 0.024 & 0.019 & 0.019 & 0.019 & 0.030 & 0.022 & 0.029 \\
 \midrule
\multirow{3}{*}{$\kappa_{13}$} & 1000 & -- & -- & -- & 0.065 & 0.056 & 0.066 & 0.124 & 0.025 & 0.099 \\
 & 2000 & -- & -- & -- & 0.040 & 0.036 & 0.039 & 0.084 & 0.017 & 0.066 \\
 & 3000 & -- & -- & -- & 0.031 & 0.030 & 0.030 & 0.089 & 0.014 & 0.065 \\
  \midrule
 \multirow{3}{*}{$\kappa_{14}$} & 1000 & -- & -- & -- & -- & -- & -- & 0.090 & 0.046 & 0.078 \\
 & 2000 & -- & -- & -- & -- & -- & -- & 0.053 & 0.031 & 0.046 \\
 & 3000 & -- & -- & -- & -- & -- & -- & 0.056 & 0.028 & 0.048 \\
  \midrule
 \multirow{3}{*}{$\kappa_{21}$} & 1000 & 0.023 & 0.022 & 0.022 & 0.028 & 0.024 & 0.029 & 0.055 & 0.030 & 0.039 \\
 & 2000 & 0.018 & 0.017 & 0.017 & 0.020 & 0.019 & 0.021 & 0.025 & 0.021 & 0.026 \\
 & 3000 & 0.014 & 0.014 & 0.014 & 0.016 & 0.015 & 0.017 & 0.024 & 0.018 & 0.023 \\
\midrule
\multirow{3}{*}{$\kappa_{22}$} & 1000 & 0.020 & 0.018 & 0.018 & 0.032 & 0.014 & 0.035 & 0.095 & 0.018 & 0.074 \\
 & 2000 & 0.016 & 0.015 & 0.015 & 0.009 & 0.009 & 0.009 & 0.041 & 0.011 & 0.014 \\
 & 3000 & 0.012 & 0.010 & 0.010 & 0.008 & 0.007 & 0.008 & 0.014 & 0.010 & 0.012 \\
 \midrule
\multirow{3}{*}{$\kappa_{23}$} & 1000 & -- & -- & -- & 0.069 & 0.054 & 0.076 & 0.098 & 0.009 & 0.117 \\
 & 2000 & -- & -- & -- & 0.038 & 0.038 & 0.039 & 0.117 & 0.006 & 0.104 \\
 & 3000 & -- & -- & -- & 0.031 & 0.030 & 0.029 & 0.098 & 0.005 & 0.083 \\
  \midrule
 \multirow{3}{*}{$\kappa_{24}$} & 1000 & -- & -- & -- & -- & -- & -- & 0.128 & 0.050 & 0.114 \\
 & 2000 & -- & -- & -- & -- & -- & -- & 0.083 & 0.034 & 0.068 \\
 & 3000 & -- & -- & -- & -- & -- & -- & 0.088 & 0.027 & 0.070\\
  \midrule
 \multirow{3}{*}{$\rho_{1}$} & 1000 & 0.018 & 0.018 & 0.018 & 0.022 & 0.024 & 0.022 & 0.044 & 0.027 & 0.044 \\
 & 2000 & 0.013 & 0.013 & 0.013 & 0.016 & 0.017 & 0.016 & 0.032 & 0.019 & 0.030 \\
 & 3000 & 0.011 & 0.010 & 0.010 & 0.014 & 0.013 & 0.015 & 0.033 & 0.015 & 0.026 \\
\midrule
\multirow{3}{*}{$\rho_{2}$} & 1000 & 0.054 & 0.055 & 0.057 & 0.044 & 0.047 & 0.045 & 0.083 & 0.055 & 0.080 \\
 & 2000 & 0.041 & 0.041 & 0.042 & 0.032 & 0.033 & 0.032 & 0.049 & 0.043 & 0.045 \\
 & 3000 & 0.036 & 0.032 & 0.036 & 0.029 & 0.027 & 0.028 & 0.042 & 0.040 & 0.037 \\
 \midrule
\multirow{3}{*}{$\rho_{3}$} & 1000 & -- & -- & -- & 0.088 & 0.055 & 0.087 & 0.080 & 0.050 & 0.074 \\
 & 2000 & -- & -- & -- & 0.037 & 0.037 & 0.038 & 0.042 & 0.029 & 0.030 \\
 & 3000 & -- & -- & -- & 0.030 & 0.029 & 0.030 & 0.049 & 0.023 & 0.034 \\
  \midrule
 \multirow{3}{*}{$\rho_{4}$} & 1000 & -- & -- & -- & -- & -- & -- & 0.094 & 0.045 & 0.087 \\
 & 2000 & -- & -- & -- & -- & -- & -- & 0.062 & 0.032 & 0.054 \\
 & 3000 & -- & -- & -- & -- & -- & -- & 0.063 & 0.026 & 0.050 \\
\bottomrule
\end{tabular}
\caption{Results of the simulation study: RMSE for $\bm{\theta}$ in each considered scenario.}
\label{tab:tab3}
\end{table}

\begin{table}[ht]
\centering
\begin{tabular}{lcccc|ccc|ccc}
\toprule%
& $K$ & \multicolumn{3}{c}{$2$} & \multicolumn{3}{c}{$3$} &
\multicolumn{3}{c}{$4$} \\
\cline{2-5}\cline{6-8}\cline{9-11}%
& $T$ & $\delta = 0.5$ & $\delta = 1$ & $\delta = 1.5$ & $\delta = 0.5$ & $\delta = 1$ & $\delta = 1.5$ & $\delta = 0.5$ & $\delta = 1$ & $\delta = 1.5$  \\
\midrule
\multirow{3}{*}{$\beta_{01}$} & 1000 & 2.031 & 1.336 & 1.332 & 2.353 & 1.682 & 1.776 & 4.262 & 1.952 & 3.847 \\
 & 2000 & 1.441 & 0.966 & 0.965 & 1.546 & 1.154 & 1.069 & 3.022 & 1.319 & 2.759 \\
 & 3000 & 1.182 & 0.820 & 0.819 & 1.049 & 0.812 & 0.806 & 3.107 & 1.040 & 2.219 \\
\midrule
\multirow{3}{*}{$\beta_{02}$} & 1000 & 0.626 & 0.543 & 0.539 & 2.144 & 0.853 & 0.926 & 3.010 & 1.738 & 2.174 \\
 & 2000 & 0.389 & 0.333 & 0.331 & 1.253 & 0.531 & 0.496 & 1.674 & 0.905 & 1.068 \\
 & 3000 & 0.312 & 0.273 & 0.275 & 0.984 & 0.399 & 0.389 & 1.511 & 0.687 & 0.814 \\
 \midrule
\multirow{3}{*}{$\beta_{03}$} & 1000 & -- & -- & -- & 1.718 & 1.627 & 1.682 & 1.294 & 0.925 & 1.288 \\
 & 2000 & -- & -- & -- & 0.736 & 0.885 & 0.775 & 0.924 & 0.559 & 0.760 \\
 & 3000 & -- & -- & -- & 0.552 & 0.499 & 0.543 & 0.758 & 0.418 & 0.627 \\
  \midrule
 \multirow{3}{*}{$\beta_{04}$} & 1000 & -- & -- & -- & -- & -- & -- & 1.204 & 0.577 & 0.925 \\
 & 2000 & -- & -- & -- & -- & -- & -- & 0.597 & 0.343 & 0.539 \\
 & 3000 & -- & -- & -- & -- & -- & -- & 0.549 & 0.288 & 0.466 \\
  \midrule
 \multirow{3}{*}{$\beta_{11}$} & 1000 & 0.235 & 0.082 & 0.082 & 0.235 & 0.127 & 0.146 & 0.556 & 0.157 & 0.355 \\
 & 2000 & 0.191 & 0.063 & 0.063 & 0.152 & 0.090 & 0.082 & 0.261 & 0.096 & 0.166 \\
 & 3000 & 0.163 & 0.048 & 0.048 & 0.110 & 0.062 & 0.059 & 0.208 & 0.084 & 0.128 \\
\midrule
\multirow{3}{*}{$\beta_{12}$} & 1000 & 0.058 & 0.056 & 0.055 & 0.215 & 0.051 & 0.051 & 0.441 & 0.145 & 0.302 \\
 & 2000 & 0.038 & 0.031 & 0.031 & 0.141 & 0.033 & 0.028 & 0.166 & 0.071 & 0.100 \\
 & 3000 & 0.031 & 0.026 & 0.026 & 0.104 & 0.022 & 0.023 & 0.134 & 0.053 & 0.134 \\
 \midrule
\multirow{3}{*}{$\beta_{13}$} & 1000 & -- & -- & -- & 0.199 & 0.299 & 0.183 & 0.137 & 0.077 & 0.111 \\
 & 2000 & -- & -- & -- & 0.073 & 0.092 & 0.069 & 0.076 & 0.030 & 0.041 \\
 & 3000 & -- & -- & -- & 0.056 & 0.054 & 0.053 & 0.045 & 0.025 & 0.036 \\
  \midrule
 \multirow{3}{*}{$\beta_{14}$} & 1000 & -- & -- & -- & -- & -- & -- & 0.179 & 0.150 & 0.181 \\
 & 2000 & -- & -- & -- & -- & -- & -- & 0.092 & 0.077 & 0.090 \\
 & 3000 & -- & -- & -- & -- & -- & -- & 0.085 & 0.063 & 0.089 \\
  \midrule
 \multirow{3}{*}{$\beta_{21}$} & 1000 & 0.133 & 0.117 & 0.118 & 0.161 & 0.135 & 0.175 & 0.330 & 0.219 & 0.431 \\
 & 2000 & 0.097 & 0.081 & 0.081 & 0.094 & 0.108 & 0.100 & 0.326 & 0.131 & 0.207 \\
 & 3000 & 0.099 & 0.066 & 0.066 & 0.079 & 0.080 & 0.079 & 0.153 & 0.091 & 0.115 \\
\midrule
\multirow{3}{*}{$\beta_{22}$} & 1000 & 0.133 & 0.122 & 0.122 & 0.105 & 0.103 & 0.117 & 0.179 & 0.146 & 0.161 \\
 & 2000 & 0.074 & 0.072 & 0.072 & 0.064 & 0.066 & 0.066 & 0.097 & 0.091 & 0.093 \\
 & 3000 & 0.061 & 0.059 & 0.059 & 0.056 & 0.058 & 0.057 & 0.075 & 0.068 & 0.073 \\
 \midrule
\multirow{3}{*}{$\beta_{23}$} & 1000 & -- & -- & -- & 0.647 & 0.643 & 0.521 & 0.230 & 0.182 & 0.232 \\
 & 2000 & -- & -- & -- & 0.194 & 0.209 & 0.203 & 0.156 & 0.110 & 0.142 \\
 & 3000 & -- & -- & -- & 0.140 & 0.135 & 0.140 & 0.139 & 0.082 & 0.117 \\
  \midrule
 \multirow{3}{*}{$\beta_{24}$} & 1000 & -- & -- & -- & -- & -- & -- & 0.235 & 0.106 & 0.186 \\
 & 2000 & -- & -- & -- & -- & -- & -- & 0.141 & 0.065 & 0.124 \\
 & 3000 & -- & -- & -- & -- & -- & -- & 0.178 & 0.049 & 0.105 \\
\bottomrule
\end{tabular}
\caption{Results of the simulation study: RMSE for $\bm{\beta}$ in each considered scenario.}
\label{tab:tab2}
\end{table}

\begin{table}[ht]
\centering
\begin{tabular}{lcccc|ccc}
\toprule%
& $K$ & \multicolumn{3}{c}{$3$} &
\multicolumn{3}{c}{$4$} \\
\cline{2-5}\cline{6-8}%
& $T$ & $\delta = 0.5$ & $\delta = 1$ & $\delta = 1.5$ & $\delta = 0.5$ & $\delta = 1$ & $\delta = 1.5$  \\
\midrule
\multirow{3}{*}{$\omega_{12}$} & 1000  & 0.091 & 0.072 & 0.086 & 0.181 & 0.127 & 0.166 \\
 & 2000 & 0.053 & 0.054 & 0.052 & 0.112 & 0.078 & 0.092 \\
 & 3000 & 0.044 & 0.045 & 0.041 & 0.114 & 0.071 & 0.091 \\
\midrule
\multirow{3}{*}{$\omega_{13}$} & 1000 & 0.148 & 0.137 & 0.148 & 0.157 & 0.110 & 0.161 \\
 & 2000 & 0.093 & 0.088 & 0.091 & 0.120 & 0.060 & 0.109 \\
 & 3000 & 0.072 & 0.072 & 0.072 & 0.113 & 0.051 & 0.103 \\
 \midrule
 \multirow{3}{*}{$\omega_{14}$} & 1000 & -- & -- & -- & 0.159 & 0.114 & 0.150 \\
 & 2000 & -- & -- & -- & 0.105 & 0.066 & 0.097 \\
 & 3000 & -- & -- & -- & 0.110 & 0.053 & 0.091 \\
  \midrule
 \multirow{3}{*}{$\omega_{21}$} & 1000 & 0.109 & 0.119 & 0.104 &  0.169 & 0.122 & 0.159 \\
 & 2000 & 0.067 & 0.080 & 0.066 & 0.100 & 0.062 & 0.096 \\
 & 3000 & 0.063 & 0.059 & 0.061 & 0.095 & 0.056 & 0.080 \\
\midrule
\multirow{3}{*}{$\omega_{23}$} & 1000 & 0.148 & 0.137 & 0.148 & 0.122 & 0.098 & 0.114 \\
 & 2000 & 0.093 & 0.088 & 0.091 & 0.072 & 0.068 & 0.070 \\
 & 3000 & 0.072 & 0.072 & 0.072 & 0.075 & 0.051 & 0.057 \\
 \midrule
\multirow{3}{*}{$\omega_{24}$} & 1000 & -- & -- & -- &  0.128 & 0.093 & 0.119 \\
 & 2000 & -- & -- & -- & 0.077 & 0.053 & 0.060 \\
 & 3000 & -- & -- & -- & 0.082 & 0.051 & 0.076 \\
  \midrule
 \multirow{3}{*}{$\omega_{31}$} & 1000 & 0.109 & 0.119 & 0.104 & 0.162 & 0.123 & 0.162 \\
 & 2000 & 0.067 & 0.080 & 0.066 & 0.133 & 0.072 & 0.123 \\
 & 3000 & 0.063 & 0.059 & 0.061 & 0.120 & 0.063 & 0.106 \\
  \midrule
 \multirow{3}{*}{$\omega_{32}$} & 1000 & 0.091 & 0.072 & 0.086 & 0.148 & 0.100 & 0.130 \\
 & 2000 & 0.053 & 0.054 & 0.052 & 0.083 & 0.067 & 0.076 \\
 & 3000 & 0.044 & 0.045 & 0.041 & 0.093 & 0.057 & 0.063 \\
\midrule
\multirow{3}{*}{$\omega_{34}$} & 1000 & -- & -- & -- & 0.162 & 0.118 & 0.146 \\
 & 2000 & -- & -- & -- & 0.099 & 0.073 & 0.091 \\
 & 3000 & -- & -- & -- & 0.102 & 0.059 & 0.084 \\
 \midrule
\multirow{3}{*}{$\omega_{41}$} & 1000 & -- & -- & -- & 0.211 & 0.166 & 0.204 \\
 & 2000 & -- & -- & -- & 0.138 & 0.089 & 0.131 \\
 & 3000 & -- & -- & -- & 0.132 & 0.075 & 0.116 \\
  \midrule
 \multirow{3}{*}{$\omega_{42}$} & 1000 & -- & -- & -- & 0.155 & 0.103 & 0.138 \\
 & 2000 & -- & -- & -- & 0.100 & 0.066 & 0.075 \\
 & 3000 & -- & -- & -- & 0.094 & 0.061 & 0.087 \\
 \midrule
 \multirow{3}{*}{$\omega_{43}$} & 1000 & -- & -- & -- & 0.172 & 0.133 & 0.164 \\
 & 2000 & -- & -- & -- & 0.119 & 0.085 & 0.106 \\
 & 3000 & -- & -- & -- & 0.113 & 0.070 & 0.096 \\
\bottomrule
\end{tabular}
\caption{Results of the simulation study: RMSE for $\bm{\omega}$ in each considered scenario.}
\label{tab:tab4}
\end{table}


\section{Segmenting marine conditions}\label{sec:application}
We have estimated a number of HSMMs from the data illustrated in Section \ref{sec:data}, by varying the number $K$ of components from 2 to 5, and using wind speed as a time-varying covariate. Results are obtained by setting $M=75$, therefore geometrically approximating the dwell time distribution tail after 1.5 days. After several attempts, we noticed that in this application larger values of $M$ yielded practically indistinguishable results, unnecessarily increasing the amount of needed storage memory.

\begin{figure}[ht]
    \centering
\includegraphics[scale=0.5]{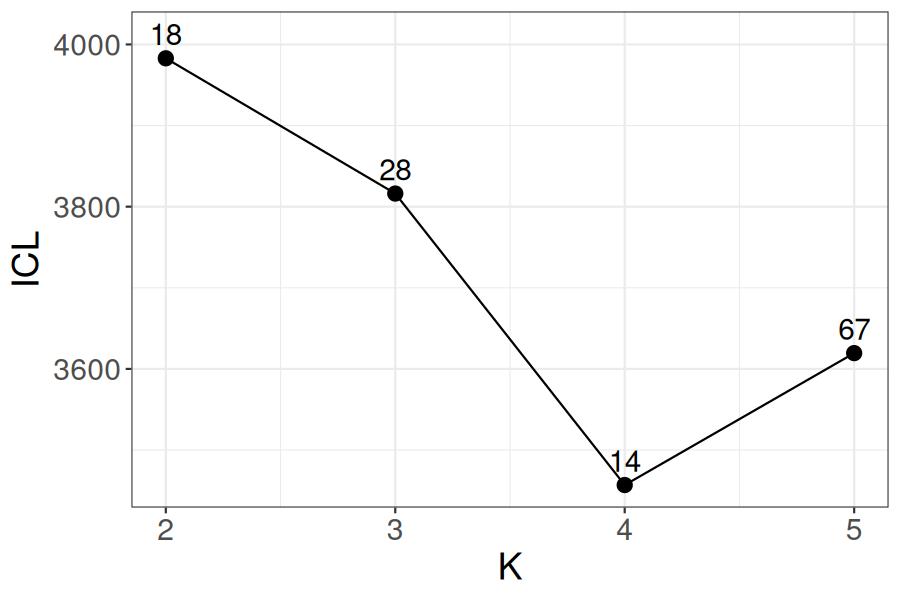}
\caption{\label{fig:icl} ICL values for a HSMM with $K=2,3,4,5$ components with related numbers of EM iterations required for convergence.}
\end{figure}

Figure \ref{fig:icl} shows, for each $K$, the number of EM iterations required to meet the convergence criterion after initialization at the best short-run parameter values and the ICL values at convergence. The drop at $K=4$ seems to indicate that a model with four components is a good compromise between goodness of fit, parsimony and latent class separation.

Table \ref{tab:parestreal} displays the estimates under this model, along with bootstrap standard errors,
computed by simulating 1000 samples, and here used to test whether estimates are significantly different of zero with a significance level set at 0.05. Some general findings in this table support some aspects of the proposed model. First, the dependence parameter $\rho$ is always significantly different of zero, supporting the choice of a toroidal distribution that accounts for circular correlation within latent regimes. The choice of using two univariate, conditionally independent circular densities, as it is often done, would have been a dispensable model restriction in this case study. Second, the regression coefficient  $\beta_2$ of wind speed is significant under two regimes (states 2 and 4). This motivates wind speed as a relevant covariate in this study and indicates that a homogeneous version of our model would have been an unnecessary shortcoming. Furthermore, being wind speed a time-varying covariate, the significance of $\beta_2$ favours a HSMM with time-varying dwell-time hazards, against a HMM with constant hazards. As an aside, the non-significance of $\beta_1$ across regimes seems to indicate that wind speed is to be held responsible for the temporal variation of the hazards.

\begin{figure}[ht]
    \centering
    \begin{subfigure}[b]{.45\textwidth}
    \centering
    \includegraphics[width=.9\textwidth]{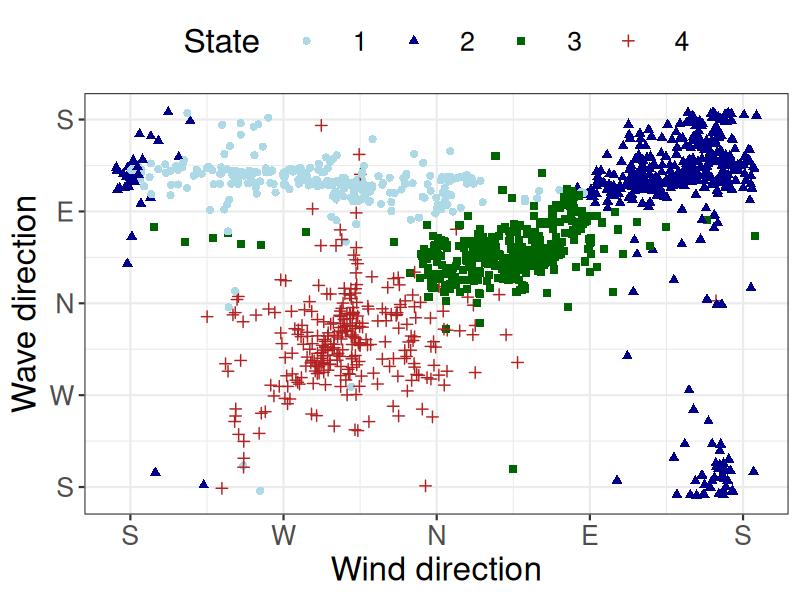}
    \caption{}
    \label{fig:classreal}
    \end{subfigure}
     \begin{subfigure}[b]{.45\textwidth}
    \centering
    \includegraphics[width=.9\textwidth]{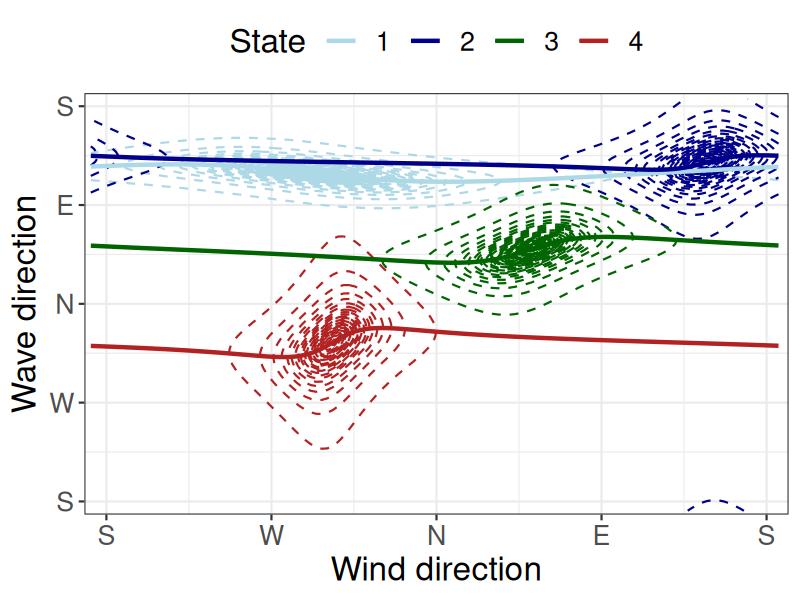}
    \caption{}
    \label{fig:contourreal}
    \end{subfigure}
     \begin{subfigure}[b]{.45\textwidth}
    \centering
    \includegraphics[width=.9\textwidth]{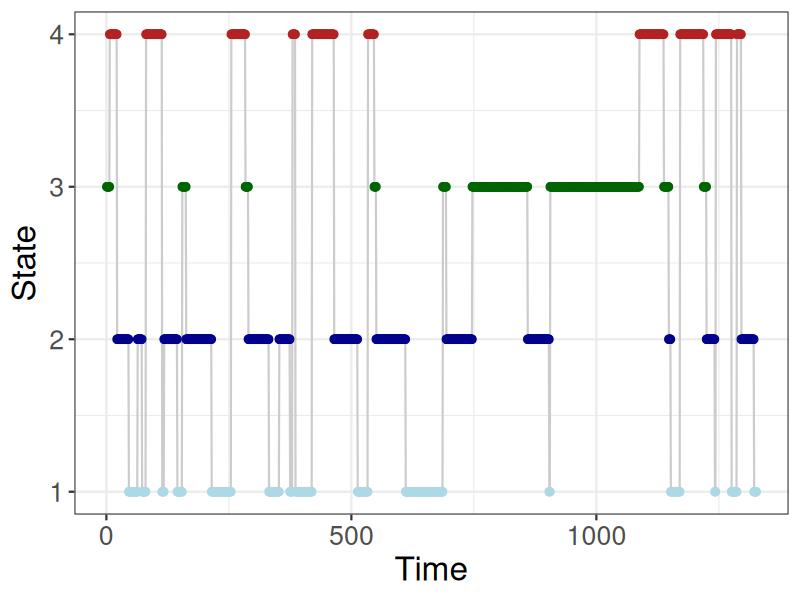}
    \caption{}
    \label{fig:chainreal}
    \end{subfigure}
     \begin{subfigure}[b]{.45\textwidth}
     \centering
     \includegraphics[width=.9\textwidth]{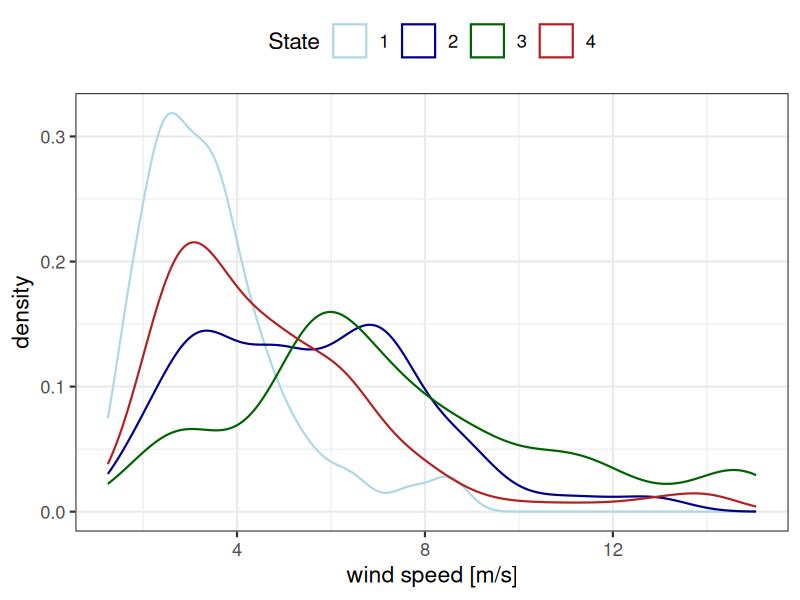}
     \caption{}
     \label{fig:conditional wind distribution}
     \end{subfigure}
    \caption{Top left: classification of the observed data (a); top right: estimated densities of the four mixture components with related circular regression lines (b); bottom left: estimated path of the hidden semi-Markov chain (c); the conditional distribution of wind speed within each latent regime (d)}
    \label{fig:resultsreal}
\end{figure}

\begin{figure}[ht]
    \centering
     \begin{subfigure}[b]{.24\textwidth}
    \centering
    \includegraphics[width=.99\textwidth]{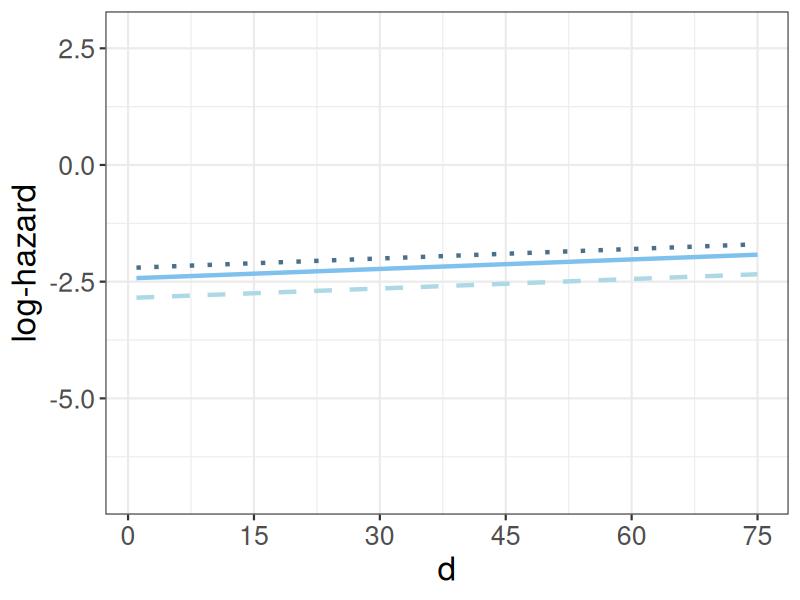}
    \caption{}
    \label{fig:hazcov1}
    \end{subfigure}
    \begin{subfigure}[b]{.24\textwidth}
    \centering
    \includegraphics[width=.99\textwidth]{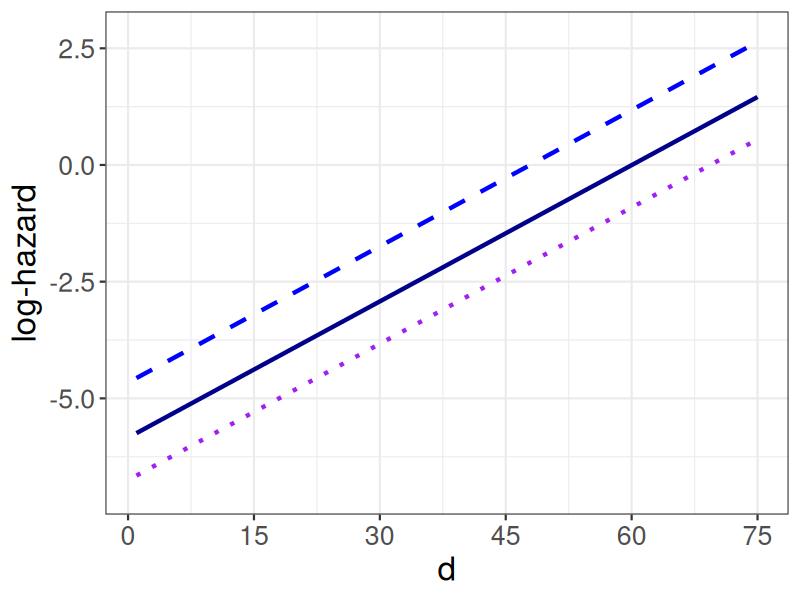}
    \caption{}
    \label{fig:hazcov2}
    \end{subfigure}
     \begin{subfigure}[b]{.24\textwidth}
    \centering
    \includegraphics[width=.99\textwidth]{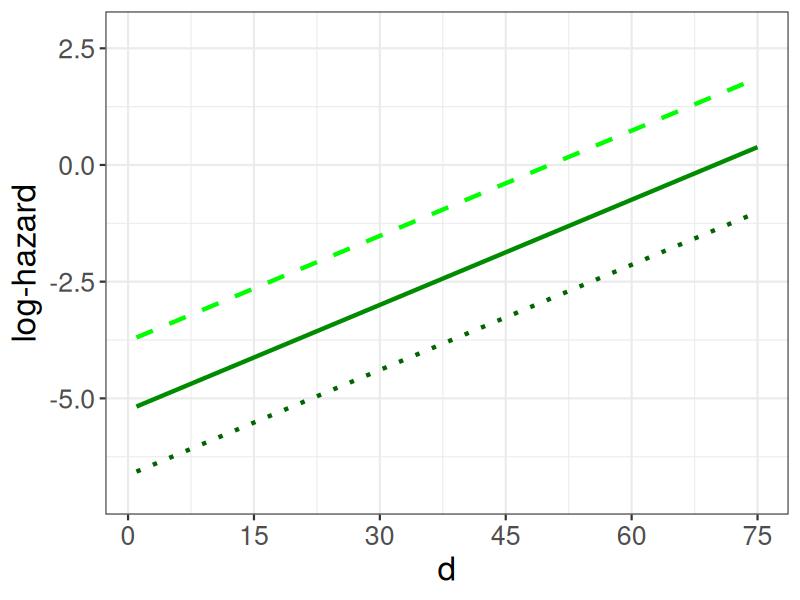}
    \caption{}
    \label{fig:hazcov3}
    \end{subfigure}
    \begin{subfigure}[b]{.24\textwidth}
    \centering
    \includegraphics[width=.99\textwidth]{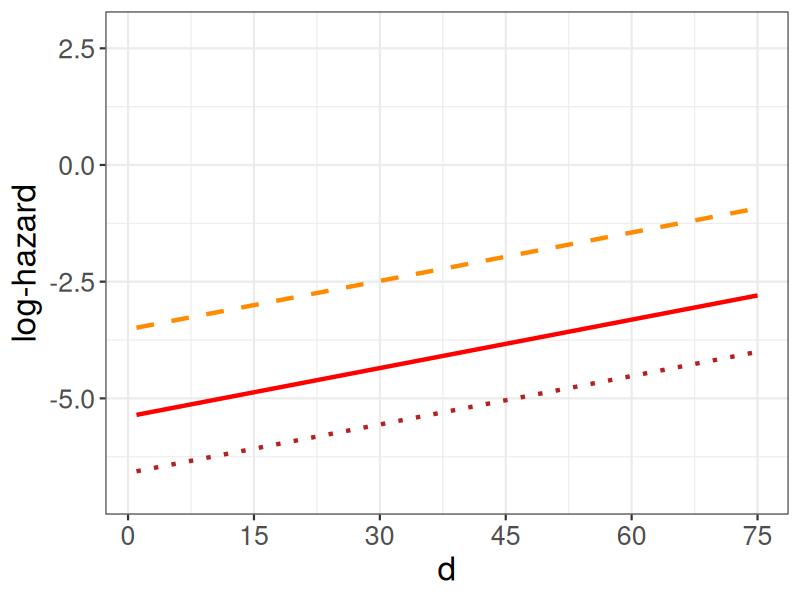}
    \caption{}
    \label{fig:hazcov4}
    \end{subfigure}
     \begin{subfigure}[b]{.24\textwidth}
    \centering
    \includegraphics[width=.99\textwidth]{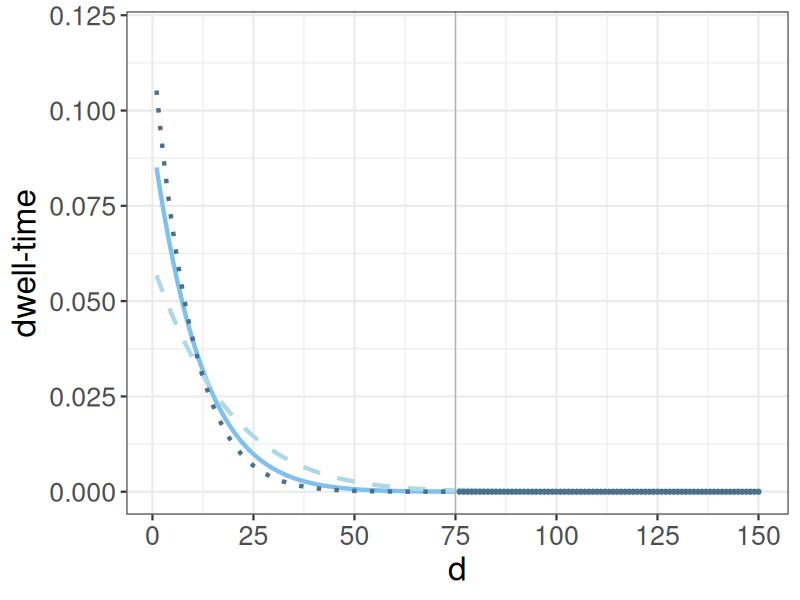}
    \caption{}
    \label{fig:dwell1}
    \end{subfigure}
    \begin{subfigure}[b]{.24\textwidth}
    \centering
    \includegraphics[width=.99\textwidth]{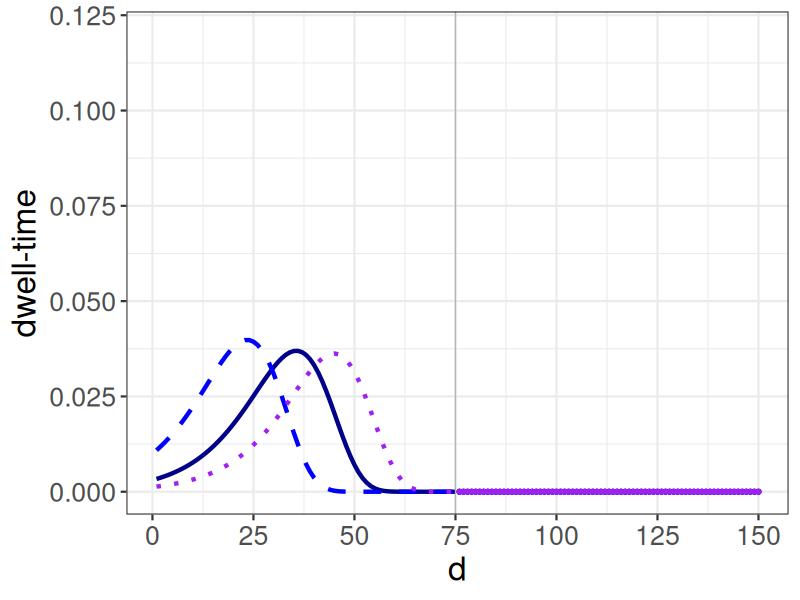}
    \caption{}
    \label{fig:dwell2}
    \end{subfigure}
     \begin{subfigure}[b]{.24\textwidth}
    \centering
    \includegraphics[width=.99\textwidth]{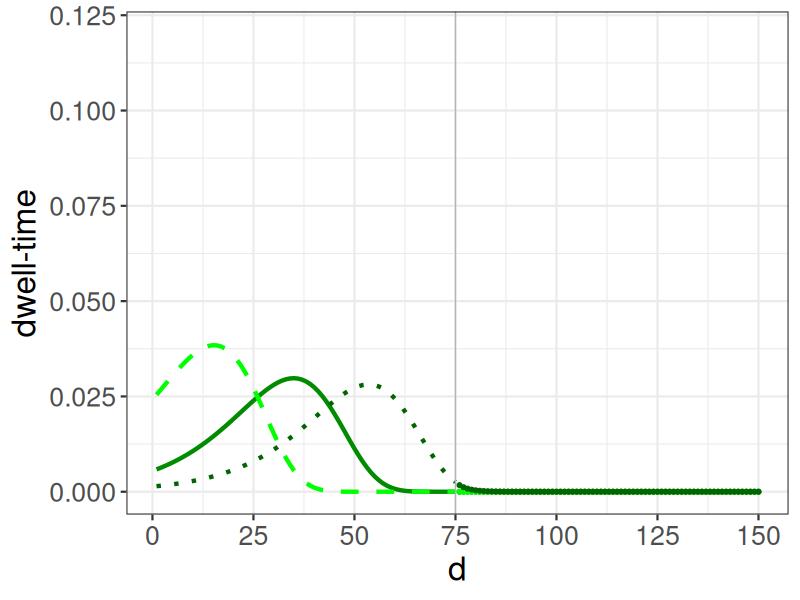}
    \caption{}
    \label{fig:dwell3}
    \end{subfigure}
    \begin{subfigure}[b]{.24\textwidth}
    \centering
    \includegraphics[width=.99\textwidth]{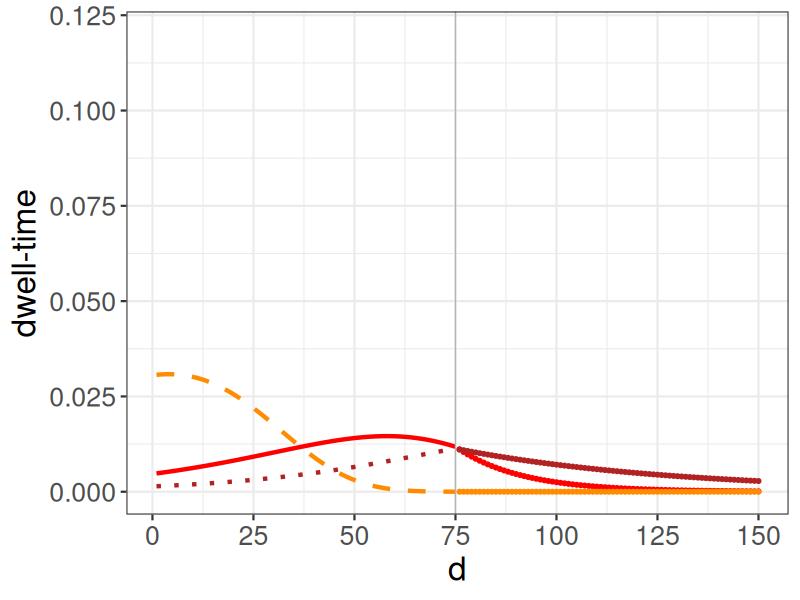}
    \caption{}
    \label{fig:dwell4}
    \end{subfigure}
\caption{(a)-(d) Hazard under three different scenarios: the covariate is fixed to its conditional first quartile (dashed), average (solid) and third quartile (dotted). (e)-(h) Corresponding dwell-time distribution in the three considered scenarios, where the points for $d>75$ show the Geometric approximation.}
    \label{fig:dwellreal}
\end{figure}

The rest of the table can be interpreted with the help of Figures \ref{fig:resultsreal} and \ref{fig:dwellreal}, which summarize the inferential output. Specifically, Figures \ref{fig:classreal} and \ref{fig:chainreal} display the data segmentation obtained in the variables space and, respectively, in the temporal domain, by associating each observation with the latent class $k$ with the highest estimated posterior probability $\hat{\pi}_{ti}$. Figure \ref{fig:contourreal} shows instead the four estimated toroidal densities in the unwrapped, Cartesian-like, toroidal space. Figure \ref{fig:conditional wind distribution} shows the (conveniently smoothed) conditional distributions of the time-varying covariate (wind speed) given each regime, obtained by associating each observation with the most probable latent class. The regime-specific dwell-time hazards and distributions displayed by Figure \ref{fig:dwellreal} were obtained by setting wind speed at the three quartiles of these conditional distributions. The vertical line at $M=75$ recalls that dwell time distributions are approximated by a geometric tail for times larger than $M=75$.

Overall, the four estimated components appear well-separated and identify four distinct sea regimes or states. Three regimes (states 2, 3 and 4) are associated with positively correlated directions, clustered around three modal directions in the northwest, the northeast and the southeast quadrant. Specifically, while states 2 and 3 occur during episodes of northern Bora winds, state 4 is the result of Sirocco episodes. This interpretation is supported by the regime-specific distributions of wind speed (Fig. \ref{fig:conditional wind distribution}). Both Bora and Sirocco are usually strong winds that drive the direction of waves. Other winds in the Adriatic Sea blow at typically moderate speed and they are not able to change the direction of the southeasterly waves that travel towards the northwest along the major axis of the Adriatic basin. This regime is captured by the first component of the model, which clusters westerly winds blowing from the Italian coast that encounter waves that travel from southeast to northwest.

The temporal segmentation obtained in Figure \ref{fig:chainreal} reflects the transition graph of the latent semi-Markov chain, as estimated by the transition probabilities of Table \ref{tab:parestreal}. Interestingly, the graph is not complete, as only states 1 and 4 are predicted to communicate with every other state. Specifically, when the process is in state 2, it is predicted to switch to state 1 (but not to state 3 and 4). Under state 2, southeastern waves are generated by a Sirocco episode and keep such direction along the major basin of the Adriatic basin as wind speed decreases (state 1). State 3 instead behaves as an intermediate state that is predicted to switch to state 2 and 4 (but not to state 1). State 3 features northeastern waves generated by a Bora episode that can change direction according to a rotation of wind toward to northwestern sector (state 4) or the southeastern sector (state 2). In general, by looking at the maximum probabilities of the estimated transition probabilities matrix, the model seems to suggest 1 $\rightarrow$ 4 $\rightarrow$ 3 $\rightarrow$ 2 $\rightarrow$ 1 as the most likely trajectory, associated with an anticyclonic atmospheric pattern.

The distinct advantage of a HSMM, compared to a more restrictive HMM, is that the temporal segmentation of the observed process (Figure \ref{fig:chainreal}) can be interpreted not only in terms of transitions between latent regimes (as it is normally done with a HMM), but also in terms of duration of each regimes. Figures \ref{fig:hazcov1}-\ref{fig:hazcov4} display the four estimated, state-specific hazards function, computed at three reference values of wind speed, namely the first quartile, the mean and third quartile of the conditional distribution of the covariate given the most probable state, as estimated by the model. Such conditional computation avoids evaluations at covariate values that rarely occur under the state of interest. Figures \ref{fig:dwell1}-\ref{fig:dwell4} display the resulting dwell time distributions, approximated by a geometric tail at dwell points greater than $M=75$. While state 1  dwells with essentially constant hazards, therefore yielding approximately geometric dwell times, states 2 and 4 feature exponentially increasing hazards that generate dwell times distributions that are clearly not geometric. For state 3, instead, we can't reject the hypothesis of time-constant hazards. According to these findings, the states associated with Bora and Sirocco episodes (2 and 4) tend to have longer durations than the states associated with intermediate regimes (1 and 3). More importantly, the durations of Bora and Sirocco episodes are positively associated with wind speed -- a result somehow expected but difficult to demonstrate without estimating a nonhomogeneous HSMM.

\begin{table}[ht]
	\centering
		\resizebox{.9\textwidth}{!}{%
		 \begin{tabular}{llcccc}
    \toprule
       & & \multicolumn{4}{c}{{{state}}}\\
\cmidrule[0.3pt](l){2-6}
        &Param. & 1 & 2 & 3 & 4 \\
        \midrule
  wind circular mean  &$\mu_1$ & -1.105 (0.134) & 2.465 (0.235) & 0.956 (0.602) & -0.943 (0.017) \\
  wave circular mean &$\mu_2$ & 1.967 (0.130) & 2.167 (0.295) & 0.887 (0.338) & -0.589 (0.149) \\
  wind circular concentration &$\kappa_1$ & 0.508 (0.064) & 0.703 (0.051) & 0.663 (0.089) & 0.762 (0.015) \\
  wave circular concentration &$\kappa_2$ & 0.847 (0.054) & 0.757 (0.020) & 0.745 (0.078) & 0.642 (0.016) \\
  wind-wave circular correlation & $\rho$ & -0.387 (0.085) & 0.183 (0.063) & 0.304 (0.115) & 0.227 (0.034) \\
  \midrule
     intercept &      $\beta_0$ & -3.766 (0.805) & -2.648 (1.32) & 0.242 (1.10) & -0.480 (1.18) \\
  time &$\beta_1$ & 0.007 (0.066) & 0.097 (0.059) & 0.075 (0.186) & 0.035 (0.081) \\
  wind speed &$\beta_2$ & 0.384 (0.400) & -0.571 (0.207) & -0.762 (0.644) & -1.019 (0.283) \\
\midrule
 && \multicolumn{4}{c}{destination state}\\
 \cmidrule[0.3pt](l){2-6}
 origin state & Param.& 1 & 2 & 3 & 4 \\
1 & $\omega_{1k}$&0  & 0.261 (0.141) & 0.231 (0.087) & 0.508 (0.116) \\
  2 & $\omega_{2k}$&0.904 (0.104) & 0  & 0.096 (0.104) & 0.000 (0.000) \\
  3 & $\omega_{3k}$&0.000 (0.000) & 0.725 (0.130) & 0  & 0.275 (0.130) \\
  4 & $\omega_{4k}$&0.309 (0.090) & 0.231 (0.069) & 0.459 (0.113) & 0 \\
         \bottomrule
    \end{tabular}
    }
    \caption{Parameters' estimates (bootstrap standard errors) of a non-homogenoeus 4-state hidden semi-Markov model.}
		\label{tab:parestreal}
\end{table}


\section{Discussion}\label{sec:discussion}
Hierarchical models offer a useful strategy to interpret complex environmental data, by allocating different features of the data to separate levels of a hierarchy. Our proposal is a hierarchical model for toroidal time series that parsimoniously integrates methods of directional statistics and survival analysis. While directional statistics is exploited at the observation level of the hierarchy by means of toroidal densities, survival analysis is employed at the latent level of the hierarchy, by using a proportional hazard model. The framework that makes such integration identifiable (and estimable) is provided by the class of the hidden semi-Markov models.

In the considered case study, the model is capable of offering a clear-cut description of wind-wave interactions in terms of intuitively appealing environmental regimes. It  provides a classification that reflects the orography of the study area, a feature often neglected by numerical models. It finally captures the influence of environmental conditions (wind speed) on the duration of specific regimes.

Tough tailored to issues that arise in marine studies, it can be easily extended to a wide range of real-world cases, where the interest is not only on the segmentation of the data according to time-varying latent classes, but also on the influence of covariates on sojourn times within each latent class.

The model is fully parametric and therefore exposed to misspecification issues. Non-parametric extensions can in principle be developed either at the observation level of the model hierarchy or at the latent level.

At the observation level, an option could be replacing the proposed mixture of toroidal densities with a non-parametric density. Methods of non-parametric estimation of toroidal densities are available \citep{dimarzio2011} and can in principle be integrated into a HSMM framework. This approach could be pursued when the interest is on the accurate estimation of the marginal data density and appropriate resources for the additional computational burden are available. However, when the interest is on segmenting the data by means of physically meaningful parameters, such as in our case study, a fully parametric strategy seems preferable and easier to perform than a non-parametric approach.

At the latent level, a non-parametric approach can be pursued by assuming a non-parametric dwell time distribution and separately estimating the whole set of mass probabilities under each regime \citep{sansom2001}. This strategy is computationally tractable only when the study involves a long time series with short dwell times and it does not guarantee against wiggly dwell time distributions with implausible gaps and spikes \citep{bulla2010}. Recently, \citet{Pohle2022} suggest penalized likelihood methods to estimate the dwell time distribution in a (homogeneous) HSMM. This idea could in principle be integrated in our proposal to estimate hazard functions non-parametrically, at the price of additional computational burden.

The influence of time-varying covariates on dwell times has been modelled by a proportional hazards regression model. During the EM run, it conveniently allows updating the dwell time regression coefficients by a weighted binomial regression model with a complementary log-log link. Alternative link functions can be used, such as the probit link, the log link, and the log-log link. It has been shown that the differences between the different models are small if the lengths of the discrete-time intervals are short compared to the study period \citep{Thompson1977}. In our case study, observations are recorded every 30 minutes, a comparatively short period with respect to the whole study duration (one month).

In general, the influence of covariates on latent dwell times can be studied by directly modelling either the probability distribution or the hazard function or the survival function, given the one-to-one correspondence between these functions. In the HSMM literature, the first approach is often preferred. In this paper, we instead pursued the second approach. By working with the hazard function, we directly test whether a time-varying covariate is responsible for a regime shift {\it given} the time spent in that regime. In other words, we are capable of identifying the conditions under which the chances of an environmental regime switch are time-constant or time-varying -- a central issue in marine studies. The third approach would open up the way to accelerated time models, which are typically understood in terms of the survival function. An idea that is certainly worth exploring in the future.

\section{Acknowledgments}
This work has been supported by MIUR, grant number 2022XRHT8R - The SMILE project: Statistical Modelling and Inference for Living the Environment.



\newpage

\end{document}